\documentclass[10pt,prb,aps,amssymb,amsmath,superscriptaddress]{revtex4}
\include{gyrokinetics-macros}
\usepackage{color,graphicx,ulem,bm}
\usepackage{bm}

\newcommand{\mbf}[1]{\mathbf{#1}}
\newcommand{\lb}{\left<}
\newcommand{\rb}{\right>}
\newcommand{\parl}{\parallel}
\newcommand{\pd}[2]{\frac{\partial #1}{\partial #2}}
\newcommand{\unit}[1]{\mathbf{\hat{#1}}}

\newcommand{\erf}{\textnormal{Erf}}
\newcommand{\secref}[1]{Sec.\ \ref{#1}}

\begin{document}

\title{Resolving velocity space dynamics in continuum gyrokinetics}

\author{M.\ Barnes}
\email{michael.barnes@physics.ox.ac.uk}
\affiliation{
Rudolf Peierls Centre for Theoretical Physics, University of Oxford, Oxford OX1 3NP, UK
}
\affiliation{
Euratom/UKAEA Fusion Association, Culham Science Centre, Abingdon OX14 3DB, UK
}
\affiliation{
Department of Physics, IREAP and CSCAMM, University of Maryland, College Park, Maryland 20742-3511, USA
}
\author{W.\ Dorland}
\affiliation{
Department of Physics, IREAP and CSCAMM, University of Maryland, College Park, Maryland 20742-3511, USA
}
\author{T. Tatsuno}
\affiliation{
Department of Physics, IREAP and CSCAMM, University of Maryland, College Park, Maryland 20742-3511, USA
}

\date{\today}

\begin{abstract}

Many plasmas of interest to the astrophysical and fusion communities are weakly collisional.  In such plasmas, small scales can develop in the distribution of particle velocities, potentially affecting observable quantities such as turbulent fluxes.  Consequently, it is necessary to monitor velocity space resolution in gyrokinetic simulations.  In this paper, we present a set of computationally efficient diagnostics for measuring velocity space resolution in gyrokinetic simulations and apply them to a range of plasma physics phenomena using the continuum gyrokinetic code \verb#GS2#.  For the cases considered here, it is found that the use of a collisionality at or below experimental values allows for the resolution of plasma dynamics with relatively few velocity space grid points.  Additionally, we describe implementation of an adaptive collision frequency which can be used to improve velocity space resolution in the collisionless regime, where results are expected to be independent of collision frequency.

\end{abstract}



\maketitle

\section{Introduction}
Velocity space dynamics are often important in the weakly collisional plasmas prevalent in astrophysics and fusion applications, leading to the necessity of a kinetic treatment.  Since the kinetic description requires a six-dimensional phase space, simulating
weakly collisional plasma processes can be computationally challenging.
Employing the gyrokinetic ordering~\cite{antonPoF80,friemanPoF82,howesApJ06} reduces the 
dimensionality by eliminating gyrophase dependence, but we are still left with a
high-dimensional system. Consequently, one would like to know how many grid points
are necessary along each dimension, particularly in velocity space, in order to resolve a 
given simulation.

In the absence of collisions or some other form of dissipation, the distribution of 
particles in velocity space can develop arbitrarily small-scale 
structures~\cite{krommesPoP94,krommesPoP99,schekPPCF08,schekApJ08}.  This
presents a problem for gyrokinetic simulations, as an arbitrarily large number of grid
points would be necessary to resolve such a system.  Of course, all physical systems
possess a finite collisionality, which sets a lower bound on the size of velocity space
structures and, therefore, an upper bound on the number of grid points required for
resolution.  We would like to know how sensitive the plasma dynamics are to the 
magnitude and form of the velocity space dissipation.  In particular, we would like
answers to the following set of questions:  Given a fixed number of grid points,
how much dissipation is necessary to ensure a resolved simulation?  Alternatively, 
given a fixed amount
of dissipation, how many grid points are necessary to ensure a resolved simulation?
Furthermore, what measurable effect, if any, does the addition of dissipation have on
collisionless plasma dynamics?

These questions have been addressed for very few plasma processes~\cite{watanabePoP04,CandyPoP06}, 
in large part due to the computational expense involved with such a study.  In this
paper, we propose computationally efficient diagnostics for monitoring velocity
space resolution, and we apply these diagnostics to a range of weakly-collisional plasma
processes using the continuum gyrokinetic code \verb#GS2#~\cite{kotschCPC95}.  
With the aid of these diagnostics, we have implemented an adaptive collision
frequency that allows us to resolve velocity space dynamics with the approximate minimal 
necessary physical dissipation.  We find that the velocity space dynamics
for growing modes are well resolved with few velocity space grid points, even in the 
collisionless limit.  Including a small amount of collisions ($\nu \ll \omega$) is necessary
and often sufficient to adequately resolve nonlinear dynamics and the long-time behavior of 
linearly damped modes.

The paper is organized as follows.  In Sec.~II we discuss velocity space dynamics 
in gyrokinetics and provide examples illustrating the development of small-scale structure
in collisionless plasmas.  Sec.~III
contains a brief overview of the \verb#GS2# velocity space grid and its dissipation mechanisms.  
We describe diagnostics for monitoring velocity space resolution in Sec.~IV and apply
them to a number of plasma processes.  In Sec.~V, we introduce an adaptive collision 
frequency and present numerical results.  We discuss our findings in Sec.~VI.

\section{Gyrokinetic velocity space dynamics}
\label{sec:gkv}

\verb#GS2# solves the coupled system consisting of the low-frequency Maxwell's equations and
the nonlinear, electromagnetic gyrokinetic equation with a model Fokker-Planck 
collision operator:
\begin{equation}
\pd{ h }{t}
+ \overbrace{\left(v_{\parl}\unit{b}+ \mbf{v_{\chi}} +\mbf{v_{B}}\right) \cdot \nabla h}^{\mathcal{K}} 
= \lb C[h] \rb_{\mbf{R}}
+ \underbrace{\frac{q F_{0}}{T}\pd{\lb \chi \rb_{\mbf{R}}}{t} 
- \mbf{v_{\chi}} \cdot \nabla F_{0}}_{\mathcal{S}},
\label{eqn:gke}
\end{equation}
where
\begin{equation}
h = f_{1} + \frac{q \Phi}{T}F_{M}
\end{equation}
is the non-Boltzmann part of the perturbed distribution function,
\begin{equation}
\mbf{v_{B}} = \frac{\unit{b}}{\Omega} \times \left[ v_{\parl}^{2} \ \left(\unit{b}\cdot\nabla\right) \unit{b}
+\frac{v_{\perp}^{2}}{2}\frac{\nabla B}{B}\right]
\end{equation}
is the sum of the curvature and $\nabla B$ drift velocities,
\begin{equation}
\mbf{v_{\chi}} = \frac{c}{B_{0}}\unit{b} \times \nabla \lb\chi\rb_{\mbf{R}}
\end{equation}
is the generalized $E\times B$ velocity (including both the $E\times B$ drift and
the drift due to the motion of the perturbed magnetic field),
\begin{equation}
\chi = \Phi - \frac{\mbf{v}}{c} \cdot \mbf{A}
\label{eqn:chi}
\end{equation}
is the generalized electromagnetic potential, $\lb \cdot \rb_{\mbf{R}}$ denotes a gyro-average 
at fixed guiding center position $\mbf{R}$, and
\begin{equation}
F_{0} = F_{M}\left(1 - \frac{q\Phi}{T}\right)
\end{equation}
is the lowest order expansion of a Maxwell-Boltzmann distribution.  The
exact form of the collision operator, $C[h]$, used in \verb#GS2# is discussed briefly
in Sec.~III and described in detail in Refs.~\onlinecite{abelPoP08} 
and~\onlinecite{barnesPoP08}.

We can group the various terms in the gyrokinetic equation (\ref{eqn:gke}) into
three distinct categories:  source terms, labeled by $\mathcal{S}$, which typically
drive large-scale structures in velocity space; convection terms, labeled by
$\mathcal{K}$, which lead to phase-mixing and the development of small-scale 
structures in velocity space; and dissipation, given by the collision operator, which smoothes
the distribution function towards a shifted Maxwellian velocity distribution.  In general, the 
structure that develops from the balancing of these terms can be quite complicated.  However,
we can gain insight into how small-scale velocity structures develop by considering 
simplified collisionless systems.  

\begin{figure}
\centering
\includegraphics[height=5.0in]{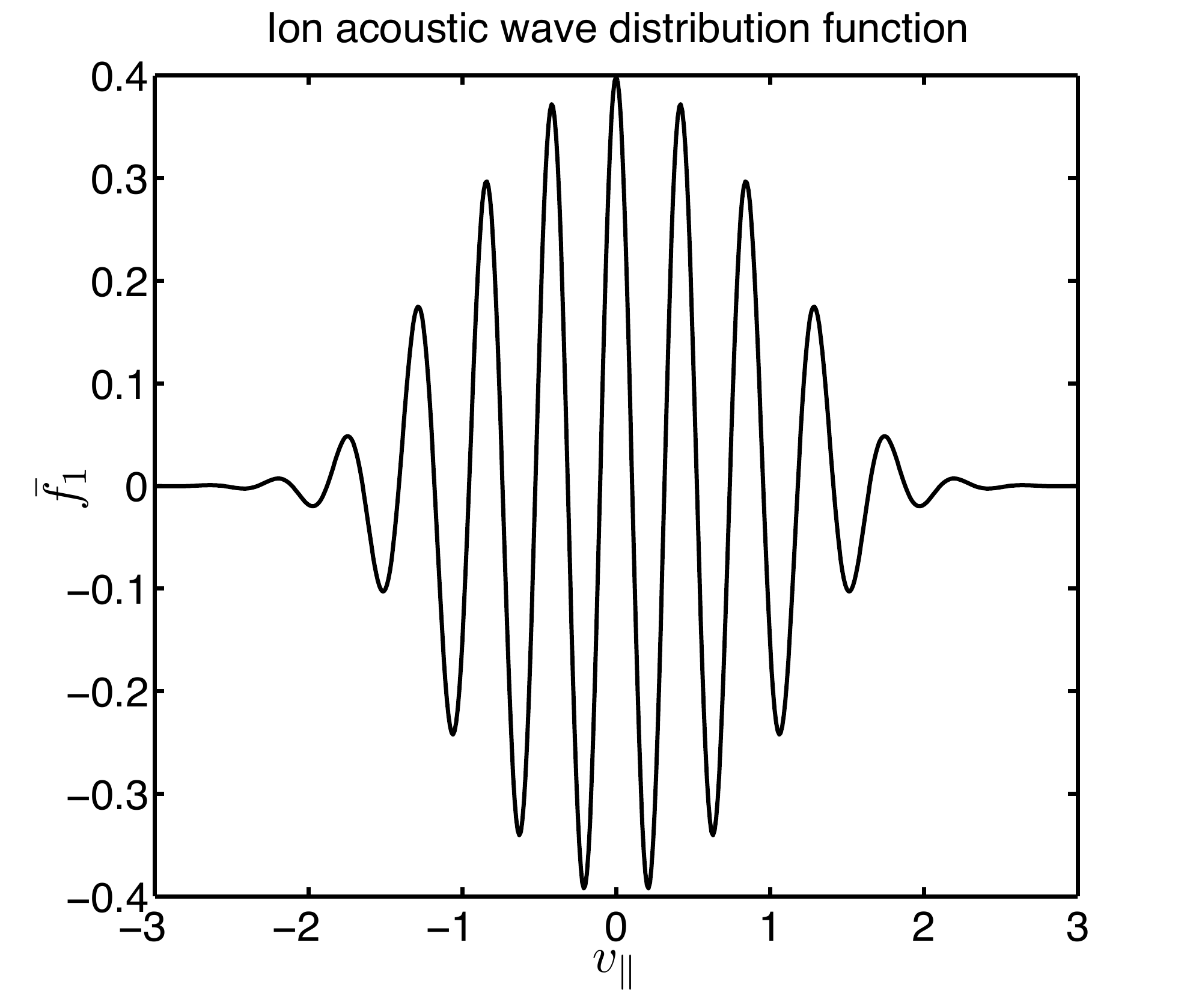}
\caption{Plot showing development of fine structure in $\bar{f}(v_{\parl})$ at $t=10 \left(k_{\parl}v_{t,i}\right)^{-1}$.  The parallel velocity on the horizontal axis is normalized by $v_{th}$, and 
$\bar{f}(v_{\parl})$ was initially a Maxwellian.}
\label{fig:iaw}
\end{figure}

In the absence of collisions, arbitrarily small scales can develop in velocity space.  This
is a result of phase-mixing, arising due to convection in real 
space~\cite{krommesPoP99,schekApJ08}.
As a simple example of this phenomenon, we include 
in Appendix A a calculation of the perturbed 
distribution function for the collisionless ion acoustic wave in a slab.  The result,
quoted here, illustrates the tendency of collisionless plasma processes to drive
small-scale velocity space structures:
\begin{equation}
\bar{f}_{1}(z,v_{\parl},t) = e^{ik_{\parl}\left(z-v_{\parl}t\right)}\mathcal{G}(v_{\parl}) + \mathcal{H}(z,v_{\parl},t),
\label{eqn:iawf}
\end{equation}
where the overbar on $f_{1}$ indicates an average over perpendicular velocities.
The quantities $\mathcal{G}$ and $\mathcal{H}$ are explicitly derived in Appendix A.  Here, it is
sufficient to note that both $\mathcal{G}$ and $\mathcal{H}$ are smooth functions of the parallel velocity.  
The presence of the oscillatory factor
 $e^{-ik_{\parl}v_{\parl}t}$ in the first term (often called the ballistic term)
leads to the development of a characteristic wavelength in velocity space that decreases 
inversely with time.
The amplitude of this ballistic term remains comparable to the second term in Eqn.~(\ref{eqn:iawf}) 
for all time, leading to the development of large amplitude oscillations
of the distribution function at arbitrarily small-scales in velocity space.  A snapshot
of this behavior at $t=10\left(k_{\parl}v_{t,i}\right)^{-1}$ is shown in Fig.~\ref{fig:iaw}.

The same calculation carried out for the collisionless ITG mode in a slab yields a 
distribution function with a similar ballistic term component.  However, since this 
mode is linearly
unstable, there is also a term describing large-scale structure in velocity space whose
amplitude grows in time to dominate the distribution function.  As a result, no 
significant small-scale structure develops.  This is a typical feature of linearly growing
modes in the collisionless limit.

Of course, all physical systems have a finite collisionality.  The dissipation arising
from this collisionality is critically important:  It is a necessary requirement for the 
existence of a statistically steady state~\cite{krommesPoP94,krommesPoP99}, and it sets a 
lower bound on the scale-size of structures in velocity space~\cite{schekApJ08}.  A simple 
estimate for the scale-size of velocity space structures can be obtained by assuming a steady 
state and balancing the collisional term with the other terms in the gyrokinetic equation.  
Noting that $C\sim\nu v_{th}^{2} \partial_{v}^{2}$ (see e.g. Refs.~\onlinecite{hirshmanPoF76} 
or~\onlinecite{cattoPoF76}), we find
\begin{equation}
\frac{\delta v}{v_{th}} \sim \sqrt{\frac{\nu}{\omega}},
\label{eqn:delvorder}
\end{equation}
where $\nu$ is the collision frequency, $\omega$ is the dynamic frequency of interest,
$v_{th}\equiv\sqrt{2T/m}$ is the thermal velocity, and $\delta v$ is the scale-size of fluctuations
in velocity space.  This estimate predicts that velocity space structures much smaller than 
the thermal velocity develop in the weakly collisional limit, $\nu \ll \omega$, as we 
would expect from our consideration of simplified collisionless systems.



\section{\texttt{GS2} velocity space}
\label{sec:gs2v}

In order to fully understand the velocity space resolution diagnostics described 
in later sections, it is necessary for the reader to have a basic knowledge of the 
way in which velocity space dynamics are treated in \texttt{GS2}.  To that purpose, 
we now give a brief explanation of the velocity space coordinates and dissipation
mechanisms employed in \texttt{GS2}.

\subsection{Velocity space coordinates}

Only two velocity space coordinates
are necessary in gyrokinetics because gyroaveraging has eliminated any gyrophase
dependence.  Fundamentally, \texttt{GS2} uses energy, $E$, and a quantity related to magnetic 
moment, $\lambda=\mu / E$, as its velocity space coordinates.  This choice
eliminates all velocity space derivatives from the collisionless gyrokinetic equation and
simplifies the discretization of derivatives in the model collision operator.  Consequently,
the spacing of the velocity space grid points is chosen to provide
accurate velocity space integrals while satisfying the necessary boundary condition
at particle bounce points.  

\subsubsection{Energy grid}

The volume element in velocity space can be written
\begin{equation}
\int d^{3}v = \frac{B_{0}}{2}\sum_{\sigma}\int_{0}^{2\pi} d\vartheta \int_{0}^{1/B_{0}}\frac{d\lambda}{\sqrt{1-\lambda B_{0}}} \int_{0}^{\infty}dv \ v^{2}
\end{equation}
where $\vartheta$ is the gyroangle and $\sigma$ denotes the sign of $v_{\parl}$.
Until recently, the energy grid in \texttt{GS2} followed the treatment of Ref.~\onlinecite{candyJCP03},
which places energy integrals in a convenient form by a change of variables to
\begin{equation}
X(x) = -\frac{2}{\sqrt{\pi}}x e^{-x^{2}} + \erf\left[x\right],
\end{equation}
where $x\equiv v / v_{th}$.
This transforms the range of integration from $x\in [0,\infty)$ to $X\in[0,1)$:
\begin{equation}
\int d^{3}v = \frac{\sqrt{\pi}}{8}B_{0} v_{th}^{3}\sum_{\sigma}\int_{0}^{2\pi} d\vartheta \int_{0}^{1/B_{0}}\frac{d\lambda}{\sqrt{1-\lambda B_{0}}} \int_{0}^{1}dX \ e^{x^{2}}.
\end{equation}
The integration domain is split into the subintervals $[0,X_{0})$ and $[X_{0},1)$,
with the perturbed distribution function assumed to be
approximately Maxwellian on $[X_{0},1)$.  
Gauss-Legendre quadrature rules~\cite{hilde} are then used to determine the location of the 
grid points in the interval $[0,X_{0})$.

\begin{figure}
\centering
\includegraphics[height=5.0in]{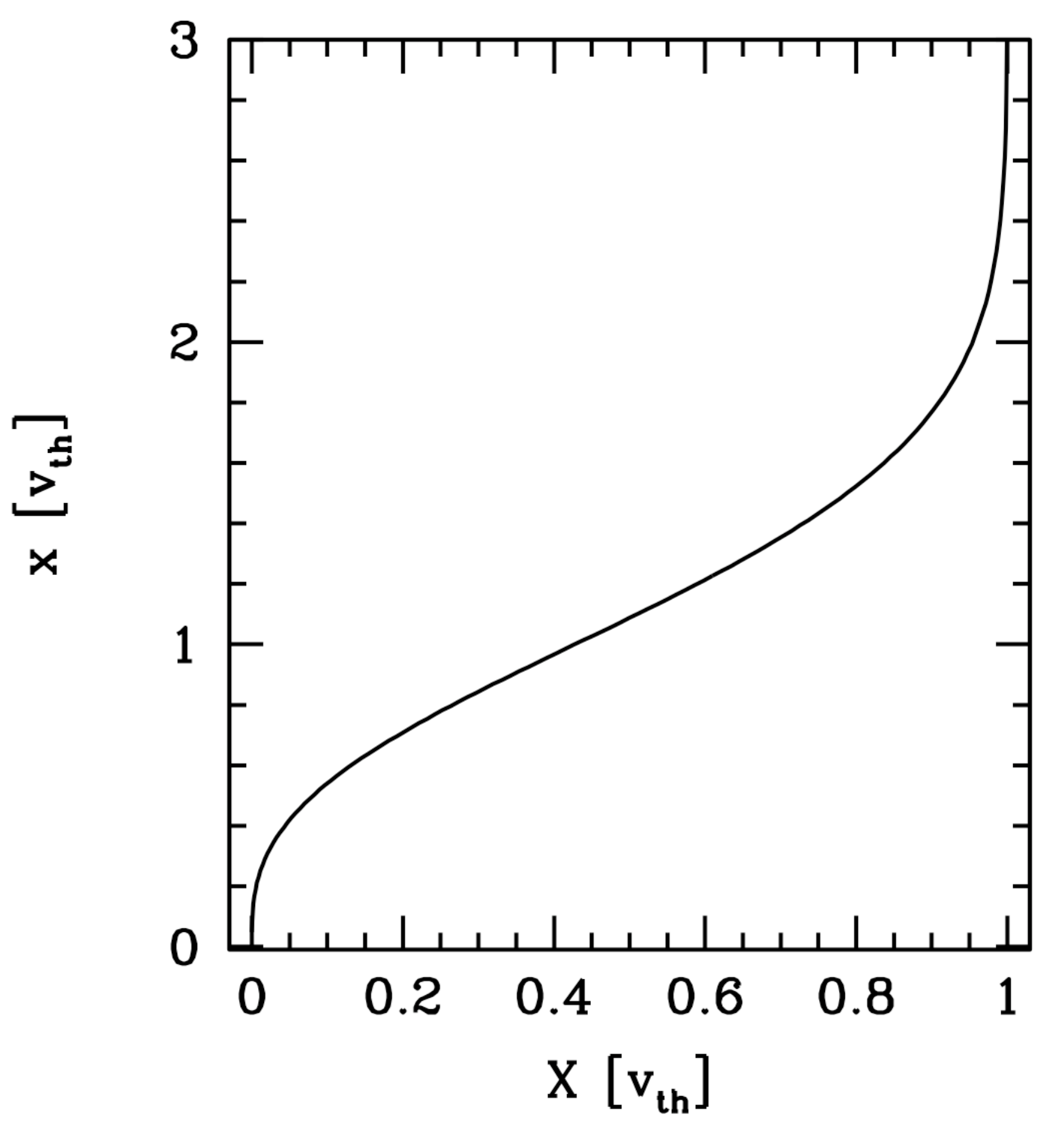}
\caption{Plot of normalized velocity $x$ over the entire $X$ domain.  The function
$x(X)$ has singularities at the boundaries of the domain due to a branch cut originating at 
$X=0$ and due to $x$ going to $\infty$ at $X=1$.}
\label{fig:xX}
\end{figure}

\begin{figure}
\centering
\includegraphics[height=5.0in]{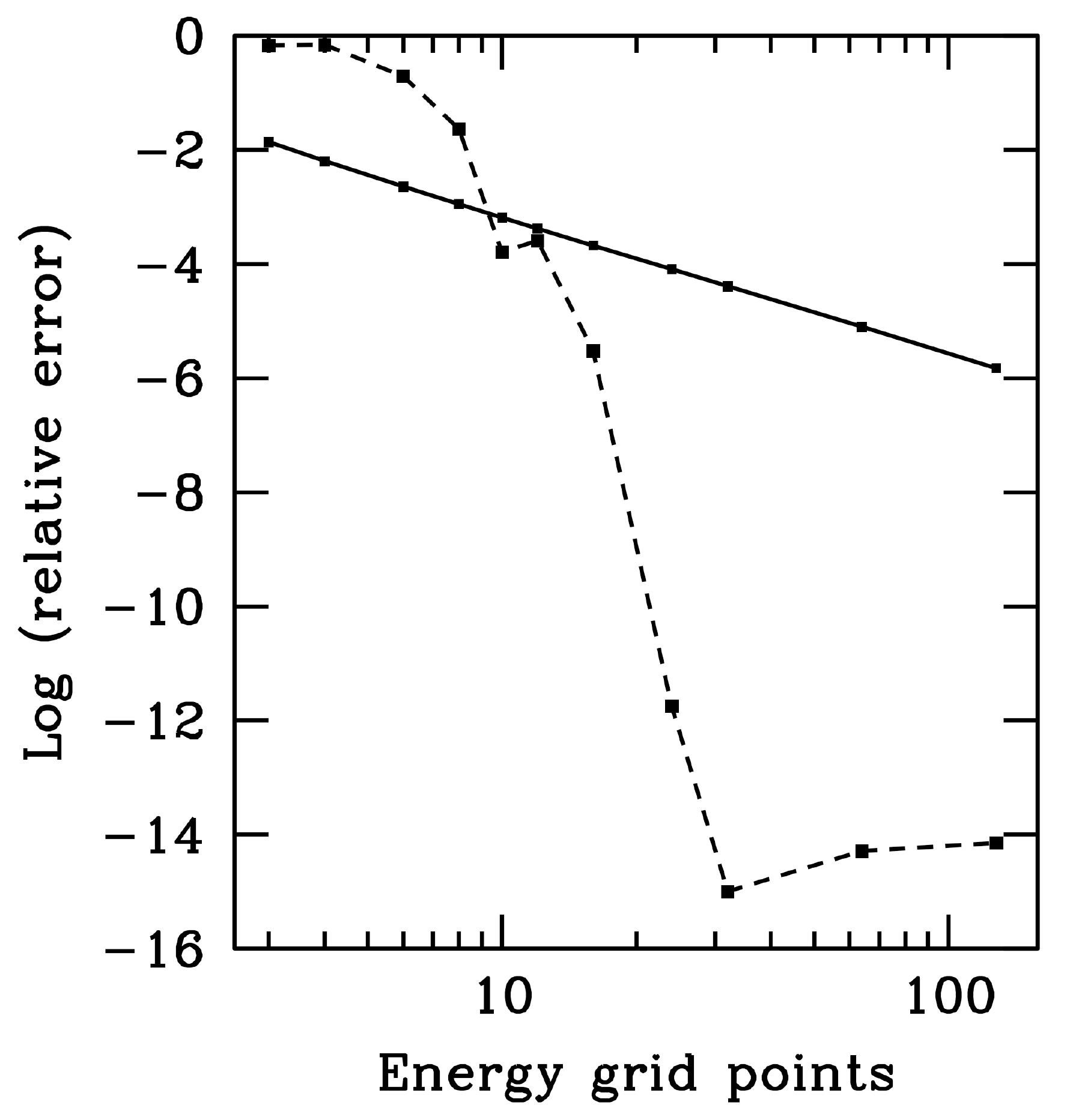}
\caption{Plot showing relative error in numerical integral of $J_{0}(k_{\perp}v_{\perp}/\Omega_{0})$ as the number of energy grid points is varied.  The error due to the integration scheme of Ref.~\onlinecite{candyJCP03} (solid line) obeys a power law in the number of grid points, with an exponent of approximately $-2.4$.  The new integration scheme detailed here (dashed line) has error approximately proportional to $(0.42N)^{-0.42N}$, where $N$ is the number of energy grid points.  Note that the minimum error in our integration scheme approaches $10^{-16}$, which is a limitation imposed by double precision evaluation of the Bessel function.}
\label{fig:j0newvold}
\end{figure}

This energy grid provides spectrally accurate energy integrals (i.e. error
$\sim (1/N)^{N}$, where $N$ is the number of energy grid points), provided the integrand
is analytic in $X$ over the integration domain (see e.g. Ref~\onlinecite{boyd}).  Unfortunately, this is seldom the 
case.  To understand why, we consider the functional form of $x(X)$.  Taylor
expanding $X$ about $x=0$, we find $X\sim x^{3}$, or equivalently, $x\sim X^{1/3}$.
This indicates a branch cut in $x$ originating from $X=0$, so that most functions of $x$ are 
non-analytic at $X=0$.  
Furthermore, one can show that $x\rightarrow\infty$ like $x\sim \sqrt{\ln \left(1/\left(1-X\right)\right)}$ as 
$X\rightarrow 1$, making $x$ non-analytic at both ends of the domain in $X$.  This can be seen in Fig.~\ref{fig:xX}, where we examine $x(X)$.  The fact that 
$x$ possesses singularities 
at the endpoints of the domain in $X$ means that the integration scheme is not
spectrally accurate for most integrands of interest (especially since the Bessel 
functions $J_{0}(k_{\perp}v_{\perp}/\Omega)$ and $J_{1}(k_{\perp}v_{\perp}/\Omega)$, 
which are non-analytic at $X=0$ and $X=1$, appear in all integrals of the
distribution function at fixed particle position $\mbf{r}$).  This is demonstrated in 
Fig.~\ref{fig:j0newvold}, where we examine the accuracy of the numerical integral 
of $h(\mbf{R})=F_{M}$ (at fixed $\mbf{r}$) as we vary the number
of velocity space grid points.

In order to achieve spectral accuracy, we have implemented a new energy grid.  We begin by
splitting the velocity integration into two separate integrals:
\begin{equation}
\int_{0}^{\infty} dx \ x^{2} G(x) = \int_{0}^{x_{0}} dx \ x^{2} G(x) + \int_{x_{0}}^{\infty} dx \ x^{2} G(x),
\end{equation}
where $x_{0}$ is a free parameter and $G(x)$ is the function we wish to integrate.  On the 
first interval, ($0, x_{0}$), we use
Gauss-Legendre quadrature rules in $x$ to obtain grid locations.  Note that use of $x$
as our integration variable ensures that the integrand $x^{2} G(x)$
will be analytic as long as $G$ is analytic in $x$ over the interval.

For the interval $(x_{0},\infty)$ we make the change of variable 
$y\equiv x^{2}-x_{0}^{2}$ to transform the integral to
\begin{equation}
\int_{x_{0}}^{\infty} dx \ x^{2} G(x)
= \frac{1}{2}\int_{0}^{\infty} dy \ e^{-y}\left[ e^{y}\sqrt{y+x_{0}^{2}} G(x)\right].
\end{equation}
We then use Gauss-Laguerre quadrature rules in $y$ to obtain grid locations.
Note that the volume element is analytic within the domain of integration, as is 
$x(y)=\sqrt{x_{0}^{2}+y}$, so that the integrand will be analytic
as long as $G$ is an analytic function of $x$.

Our use of spectral integration techniques (i.e. Gaussian quadrature), coupled with
the analyticity of our integrand for well-behaved functions $G(x)$, ensures
the spectral accuracy of our integration scheme.  While an exponential order of convergence
is assured, the rate of convergence depends on the exact nature of the integrand and our
choice of the parameter $x_{0}$.  In general we choose $x_{0}\gtrsim 2.5$ so that the branch
cut at $y=-x_{0}^{2}$ is sufficiently far from the domain of integration in $y$ to minimally
impact the rate of convergence.  We demonstrate the spectral
accuracy of the scheme and determine the rate of convergence in Fig.~\ref{fig:j0newvold}.  It is worthwhile to note that for few grid points ($\lesssim 8$ in Fig.~\ref{fig:j0newvold}) 
the grid given in Ref.~\onlinecite{candyJCP03} may be more accurate.  This is because 
the energy variable $X$
eliminates velocity-dependence of the volume element (when solving for the normalized 
distribution function $\tilde{h}\equiv h/F_{0}$), while the new v-space integrals 
described here have the velocity-dependent volume element $x^{2}e^{-x^{2}}$ that must be 
integrated regardless of the form of $\tilde{h}$.

\subsubsection{Lambda grid}

For systems with curved magnetic field lines, special care is also required when dealing with 
$\lambda$~\cite{kotschCPC95}.  There are two
reasons for this:  the grid points provided by Gaussian quadrature rules are concentrated
near the endpoints of the domain, whereas one would like them to be concentrated at
the trapped-passing boundary; and one must ensure that the proper boundary condition
(i.e. $f(v_{\parl}=0^{+})=f(v_{\parl}=0^{-})$) is satisfied at each of the bounce points.  Consequently, the $\lambda$-grid is divided
into two regions corresponding to trapped and untrapped particles, respectively.

For values of $\lambda$ such that $0 \leq \lambda < 1/B_{max}$, the
corresponding particles are untrapped by the magnetic potential well.  In this region
of velocity space, the integration variable $\tilde{\xi}\equiv\sqrt{1-\lambda B_{max}}$ is 
chosen.  It is similar to pitch-angle, but it has no spatial dependence.  Similarly to the
energy, Gauss-Legendre quadrature rules are used to obtain the location of grid points
in  $\tilde{\xi}$.  This naturally provides a concentration of gridpoints near the 
trapped-passing boundary.

\begin{figure}
\centering
\includegraphics[height=5.0in]{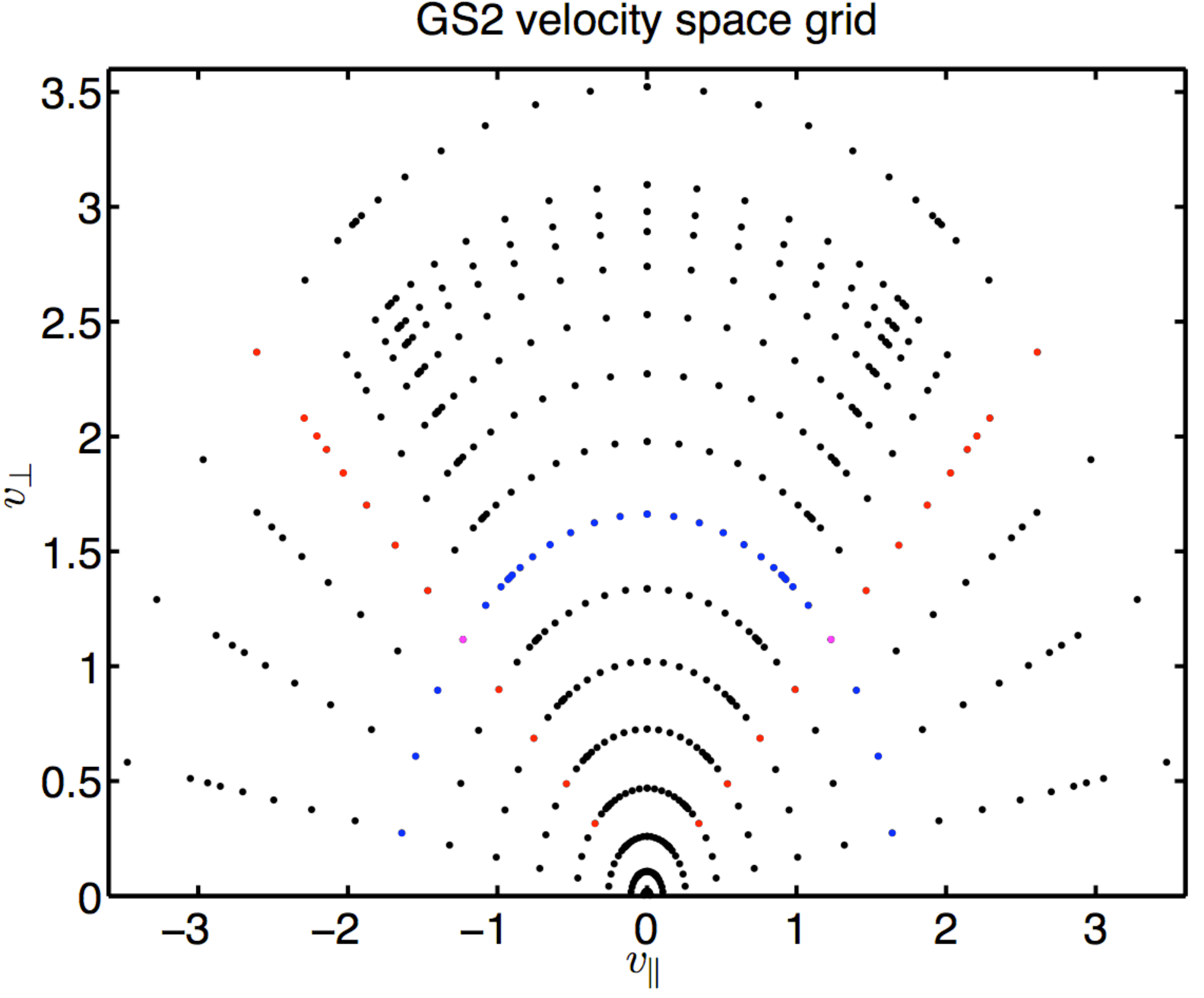}
\caption{Typical velocity space grid used in \texttt{GS2}.  Grid points are concentrated near
the trapped-passing boundary (whose location varies with $\theta$) and at lower
energy values where the Maxwellian weighting dominates.  Red (blue) grid points are sample $\lambda$ (energy) grid points that are dropped when calculating integral approximation with lower degree of precision.}
\label{fig:GS2v}
\end{figure}

For values of $\lambda$ such that $1/B_{max} < \lambda < 
1/B_{min}$, the corresponding particles are trapped by the magnetic potential well.
In the trapped region, grid points are chosen to fall on bounce points in order to
allow for the enforcement of boundary conditions.  Mathematically, this means that
for each value of $\theta$, there must be a corresponding 
$\lambda$ such that
\begin{equation}
\xi(\theta)=\frac{v_{\parl}(\theta)}{v}=\sqrt{1-\lambda B_{0}(\theta)} = 0,
\end{equation}
where $\theta$ gives the position along the unperturbed
magnetic field line and $\xi$ is the pitch-angle.  This choice of $\lambda$ values also leads 
to a concentration 
of grid points near the trapped-passing boundary.  A typical \texttt{GS2} grid layout for a system 
with trapped particles is shown in Fig.~\ref{fig:GS2v}.  It should be noted that
the $\lambda$ integrals, like the energy integrals, are spectrally accurate, 
provided the distribution function is analytic in $\xi$.

\subsection{Velocity space dissipation}

Some form of dissipation is often necessary to prevent the formation
of arbitrarily small-scale structures in velocity space.  This can be achieved either through
artificial numerical dissipation or through implementation of a model collision operator.
Both options are available in \texttt{GS2}.

\subsubsection{Model collision operator}

\texttt{GS2} uses a model Fokker-Planck collision operator that includes the effects of pitch-angle
scattering and energy diffusion while satsifying Boltzmann's H-Theorem and conserving 
particle number, momentum, and  energy~\cite{abelPoP08, barnesPoP08}:
\begin{equation}
C[h] = \mathcal{L}[h] + \mathcal{D}[h] + \mathcal{M}[h],
\label{eqn:collop}
\end{equation} 
where
\begin{equation}
\mathcal{L}[h] = \frac{\nu_{D}}{2}\left(\pd{}{\xi}\left(1-\xi^{2}\right)\pd{h}{\xi} + \frac{1}{1-\xi^{2}}\pd{^{2}h}{\vartheta^{2}}\right)
\end{equation}
is the Lorentz collision operator,
\begin{equation}
\mathcal{D}[h] = \frac{1}{4x^{2}} \pd{}{x}\left(\nu_{s}x^{2}F_{0} \pd{}{x}\frac{h}{F_{0}}\right)
\end{equation}
is the energy diffusion operator, and $\mathcal{M}[h]$
contains momentum- and energy-conserving corrections.  The velocity-dependent
collision frequencies $\nu_{s}$ and $\nu_{D}$ are given by
\begin{equation}
\nu_{s} = \frac{2\nu_{\alpha \beta}}{x^{3}}\left(\erf[x] - \frac{2xe^{-x^{2}}}{\sqrt{\pi}}\right)
\end{equation}
and
\begin{equation}
\nu_{D} = \frac{1}{x^{2}}\left(\nu_{\alpha \beta}\frac{\erf[x]}{x} - \frac{\nu_{s}}{4}\right).
\end{equation}
with $\nu_{\alpha\beta}$ the frequency of collisions of particles of species $\alpha$
with particles of species $\beta$.  A detailed description of the collision operator is
given in Ref.~\onlinecite{abelPoP08}.
Here we simply present the gyroaveraged collision operator in spectral form:
\begin{equation}
\begin{split}
\left< C[h] \right>_{k} &= \frac{\nu_{D}}{2}\pd{}{\xi}\left(1-\xi^{2}\right)\pd{h_{k}}{\xi} + \frac{v_{th}^{2}}{4v^{2}}\pd{}{v}\left(\nu_{s}v^{2}F_{0}\pd{}{v}\frac{h_{k}}{F_{0}}\right) \\
&- \frac{k_{\perp}^{2}\rho^{2}}{8\Omega_{0}^{2}}\left(\frac{2v^{2}}{v_{th}^{2}}\nu_{D}\left(1+\xi^{2}\right)+\nu_{s}\left(1-\xi^{2}\right)\right)h_{k}+\nu_{E} v^{2} J_{0}(a) F_{0} \frac{\int d^{3}v \ \nu_{E} v^{2} J_{0}(a) h_{k}}{\int d^{3}v \ \nu_{E} v^{4} F_{0}}\\
&+\nu_{D}F_{0}\Big(J_{0}(a)v_{\parl}
\frac{\int d^{3}v \ \nu_{D}v_{\parl}J_{0}(a)h_{\mbf{k}}}{\int d^{3}v \ \nu_{D}v_{\parl}^{2}F_{0}} 
+ J_{1}(a)v_{\perp}\frac{\int d^{3}v \ \nu_{D}v_{\perp}J_{1}(a)h_{\mbf{k}}}{\int d^{3}v \ \nu_{D}v_{\parl}^{2}F_{0}}\Big)\\
&-\Delta \nu F_{0}\Big(J_{0}(a)v_{\parl}
\frac{\int d^{3}v \ \Delta \nu v_{\parl}J_{0}(a)h_{\mbf{k}}}{\int d^{3}v \ \Delta \nu v_{\parl}^{2}F_{0}} 
+ J_{1}(a)v_{\perp}\frac{\int d^{3}v \ \Delta \nu v_{\perp}J_{1}(a)h_{\mbf{k}}}{\int d^{3}v \ \Delta \nu v_{\parl}^{2}F_{0}}\Big)
\end{split}
\label{gyroC}
\end{equation}
where $k$ is the perpendicular wavenumber, $a\equiv kv_{\perp}/\Omega_{0}$, $\Delta \nu=\nu_D-\nu_s$, and 
\begin{equation}
\nu_{E} = \frac{2\nu_{\alpha\beta}}{x^{3}}\left(\erf[x] - \frac{4xe^{-x^{2}}}{\sqrt{\pi}}\right).
\end{equation}
Details on numerical implementation of the collision 
operator (\ref{gyroC}) can be found in Ref.~\onlinecite{barnesPoP08}.

\subsubsection{Numerical dissipation}

Numerical dissipation enters in \texttt{GS2} through two mechanisms.  The first is the optional 
decentering of spatial and temporal finite differences, as described in Ref.~\onlinecite{kotschCPC95}.  
The lowest order contribution to dissipation due to decentering in time and space is
\begin{eqnarray}
\pd{^{2}h}{t \partial \theta}\left[\Delta \theta \left(\delta - \frac{1}{2}\right)+ \left(v_{\parl}\right)_{j+1/2}\Delta t\left(\beta - \frac{1}{2}\right)\right]
-\pd{^{2}\lb\chi\rb_{\mbf{R}}}{t \partial \theta}\left[\Delta \theta \left(\delta - \frac{1}{2}\right)\frac{qF_{0}}{T}\right],
\end{eqnarray}
where $\Delta \theta$ is the grid spacing along the field line, $\Delta t$ is the time step
size, $\delta$ is a parameter that allows for spatial upwinding (when $\delta \neq 1/2$), and $\beta$ is a parameter that allows for the variation of the
time discretization between fully explicit ($\beta=0$)
and fully implicit ($\beta=1$)~\footnote{\texttt{GS2} actually uses $\tilde{\beta}=\beta-1/2$, 
but we choose to use $\beta$ here for simplicity.}.  

\begin{figure}
\centering
\includegraphics[height=5.0in]{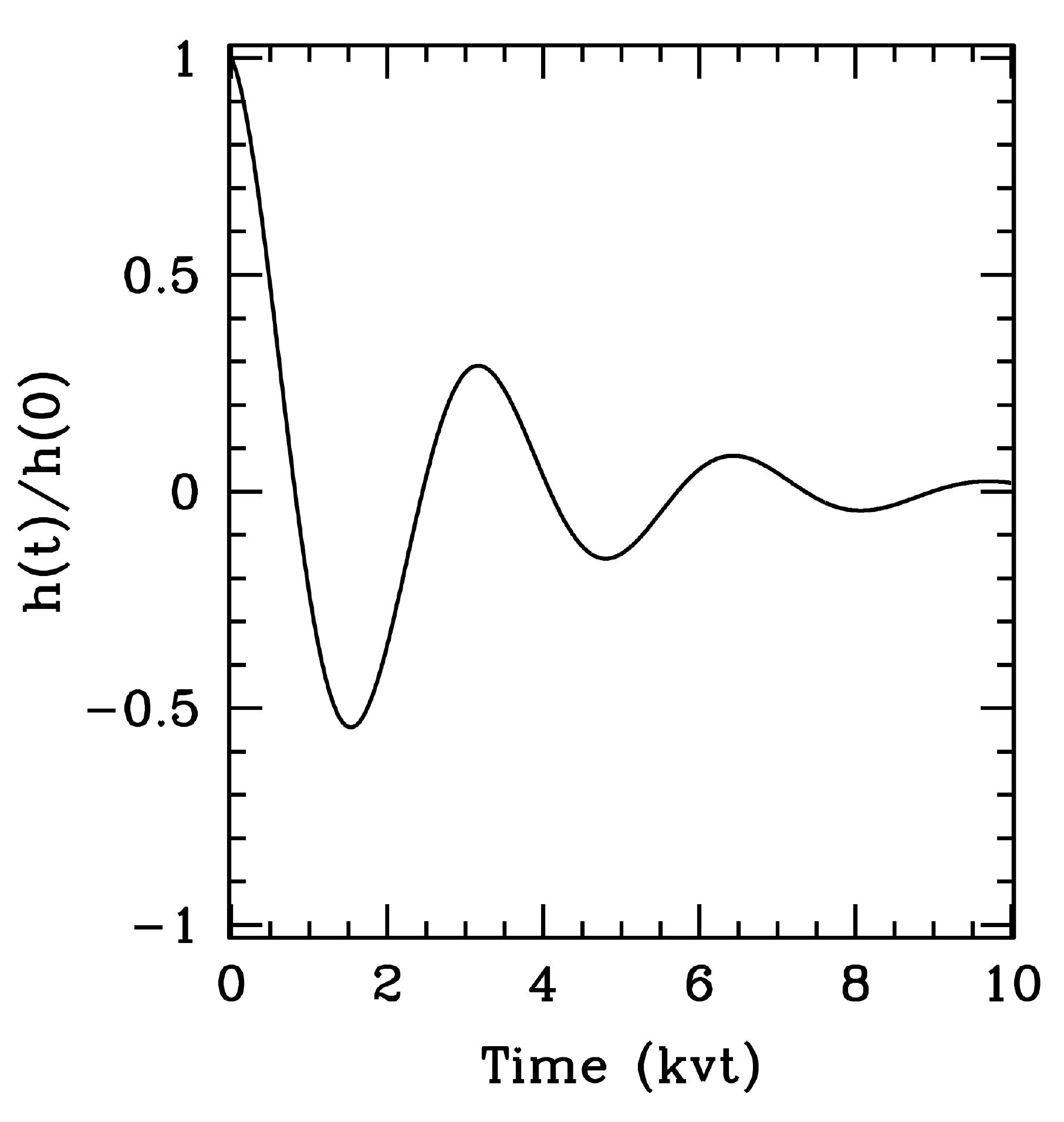}
\caption{Damping of the real part of the distribution function $h$ [Eqn.~(\ref{eqn:numdiss})]
as a result of decentered finite differences in space and time.  Here, we are considering 
$kv\Delta t=k\Delta \theta=0.2$, and $\beta=\delta=1.0$ (fully implicit, upwind).}
\label{fig:numdiss}
\end{figure}

In order to see how this term leads to
dissipation, we consider the simplified system governed by the equation
\begin{equation}
\pd{h}{t} + v\pd{h}{\theta}= 0.
\end{equation}
Finite differencing this equation using the scheme given in Ref.~\onlinecite{kotschCPC95}, 
we find that numerically we are solving the equation
\begin{equation}
\pd{h}{t} + v\pd{h}{\theta} \approx -\pd{^{2}h}{t \partial \theta}\left[\Delta \theta \left(\delta - \frac{1}{2}\right)+ v\Delta t\left(\beta - \frac{1}{2}\right)\right].
\end{equation}
Assuming $h=\tilde{h}(t) e^{ik\theta}$, we obtain the solution
\begin{equation}
\tilde{h}(t) \sim \exp\left[\frac{kvt}{i-k\left(\Delta \theta \left(\delta-1/2\right) 
+ v\Delta t\left(\beta-1/2\right)\right)}\right],
\label{eqn:numdiss}
\end{equation}
which is damped unless $\beta=\delta=1/2$, as show in Fig.~\ref{fig:numdiss}.
While decentering of finite differences can sometimes improve numerical stability, care must
be taken to ensure such artificial dissipation does not lead to unphysical behavior.  This
is typically done by monitoring the ratio of artificial to physical dissipation, which, ideally, 
should be small.

The second source of numerical
dissipation arises in systems with sheared magentic fields due to the necessity of a
'twist-and-shift' parallel boundary 
condition~\cite{beerPoP95}.  This non-periodic boundary condition couples modes at 
opposite ends of the simulation domain 
along the field line.  Since only a finite number of modes can be kept in a simulation,
some modes will eventually couple to modes that are not present, and this information is
lost.  The information that is lost is replaced by a smoothed distribution function, which should be associated with an increase in the entropy of the system.  This entropy generation should be diagnosed in order to verify that it is small compared to the entropy generated by collisions.

\section{Velocity space resolution diagnostics}
\label{sec:vresdiag}

There are numerous ways in which one could try to determine whether or not a particular
simulation is well-resolved in velocity space.  
Ideally, one would perform a grid convergence study for each simulation; if 
quantities of interest are unchanged by doubling the number of grid points, one can feel
relatively confident in the simulation results.  However, this process is computationally
expensive, as it involves running a simulation multiple times with an excessive number
of grid points.  Consequently, it is not desirable to perform a grid convergence study for every
simulation.  In practice, one tests convergence for a problem thought to be resolution intensive 
and posits that other simulations, which likely require fewer grid points, are therefore 
resolved.  Unfortunately, one seldom knows in advance how fine the structure in
velocity space will become, so one can't be fully confident that every simulation is resolved.

An alternative approach that has recently gained popularity in the computational plasma 
physics community involves monitoring entropy balance in the system~\cite{watanabePoP04,CandyPoP06}.
The entropy balance relation arises from multiplying the gyrokinetic equation (\ref{eqn:gke}) by $hT_{0}/F_{0}$ and integrating over all phase space.
Since the gyrokinetic equation itself is automatically satisfied by a gyrokinetic solver,
the only possible sources of inbalance in this relation come from numerical 
dissipation and errors in the numerical approximations to phase space integrals.  If
the change in entropy due to numerical dissipation is also diagnosed and included
in the entropy balance, as is often the case, then we are left with errors due only
to phase space integration.  Since the errors in these particular integrals are not directly
related to errors in the calculation of the distribution function at the newest timestep,
they do not necessarily correlate with the simulation resolution.  In
particular, one could easily define a poorly-resolved system for which this diagnostic
predicts perfect entropy balance.  One such example is the linear, collisionless ion acoustic 
wave in a slab (treated in detail in Appendix A).  For this case, we numerically find 
entropy balance despite the fact that the numerical damping rate goes bad due to poor 
resolution in velocity space.

Of course, one could simply produce plots or movies of the distribution function in 
velocity space over the course of the simulation to see if structure develops at the gridscale. 
This is undoubtedly useful and possibly sufficient in some cases.  However, what exactly 
one sees depends on how the data is visualized; for data on irregularly spaced grids, the
interpolation scheme used to generate the images often introduces erroneous or misleading
structure.  Furthermore, for simulations involving non-trivial spatial structure,  one would 
have to examine movies of the distribution function at each point in physical space.  
This is a memory- and time-intensive approach that is rarely feasible.

We would like to have computationally cheap diagnostics that provide real-time 
information on velocity space resolution that is easy to analyze and interpret.  
In the following subsections, 
we present two such diagnostics developed for implementation in \texttt{GS2} that could easily
be adapted for use in other continuum kinetic simulations.

\subsection{Integral error estimates}

Upon consideration of the collisionless gyrokinetic-Maxwell's system of equations, one 
finds that the only
nontrivial operation in velocity space is integration, which enters in the calculation of the
electromagnetic fields.  Consequently, resolution in velocity space is limited only by the
accuracy with which the velocity space integrals are calculated.  By calculating the error
in our numerical integration, we are thus able to monitor velocity space resolution.

In particular, when we discretize the gyrokinetic equation, we obtain an equation of the form
\begin{equation}
g_{j+1} = G\left[g_{j},\Phi_{j},\Phi_{j+1},\chi_{j},\chi_{j+1}\right],
\label{eqn:gnew}
\end{equation}
where $g\equiv\lb f_{1} \rb$ is the perturbed, guiding center distribution function evolved by \verb#GS2#, $\Phi$ is the 
electrostatic potential, $\chi$ is the generalized
electromagnetic potential defined in Eqn.~(\ref{eqn:chi}), $G$ is 
a function that depends on the details of the numerical scheme, and the subscript denotes
the timestep.  We assume that the 
time-converged solution for $g$ is independent of the initial condition.  Since
using the calculated $g_{j}$ and $\Phi_{j}$ is equivalent to specifying a new initial 
condition, we find that the time-converged solution is independent of errors in $g$ and 
$\Phi$ at earlier timesteps.  This is convenient because it means we can monitor
resolution merely by calculating the error made in the latest timesteps of a time-converged
simulation.

Ideally, we would accomplish this 
by calculating estimates for the error in $\Phi_{j+1}$ and $\chi_{j+1}$ and plugging
these into Eqn.~(\ref{eqn:gnew}) to obtain an error estimate for $g_{j+1}$.  This
might be feasible for linear systems, but the presence of nonlinear terms makes this approach
computationally prohibitive.  Consequently, we must define an alternative quantity whose 
error estimate is cheaper to compute, but that can still be used as a means
of monitoring velocity space resolution.  There are numerous possible candidates; we 
choose to compute two quantities, $v_{\Phi}$ and $v_{A}$, related to $\nabla_{\perp}\Phi$
and $\nabla_{\perp}A_{\parl}$:
\begin{equation}
 \left(\begin{array}{c} v_{\Phi}\\
v_{A}\end{array}\right) = \textnormal{max}\{k_{x},k_{y}\}  \left(\begin{array}{c} \Phi(\theta, k_{x},k_{y})\\
A_{\parl}(\theta,k_{x},k_{y})\end{array}\right),
\label{eqn:errquant}
\end{equation}
where $k_{x}$ and $k_{y}$ are the wavenumbers corresponding to the coordinates 
$x\equiv \left(\psi - \psi_{0}\right)q_{0}/B_{0}r_{0}$ and
$y\equiv-\left(\alpha - \alpha_{0}\right)r_{0}/q_{0}$~\cite{beerPoP95}.
Here, $\psi$ is the 
poloidal flux, $\alpha$ is the field line label, $B_{0}$ is the background magnetic field at 
the magnetic axis,
$r_{0}$ is the distance from the magnetic axis to the center of the simulation domain, and
$q_{0}$ is the safety factor on the field line of interest, labeled by 
$(\psi_{0},\alpha_{0})$.
The quantities in Eqn.~(\ref{eqn:errquant}) were chosen because, with the exceptions of the 
parallel convection term and one source term,
$\Phi$ and $A_{\parl}$ always enter the gyrokinetic equation for $g$ multiplied by
either $k_{x}$ or $k_{y}$.
Therefore, it is reasonable that this $k$-weighted quantity is most likely to be responsible
for errors in $g_{j+1}$.  Although not considered here, the expression (\ref{eqn:errquant})
could potentially be improved by including $k_{\parl}$ in the max operator.  This would
take into account the effect of the parallel convection term.  However, recent
theoretical~\cite{schekApJ08} and numerical~\cite{tatsunoPRL08} work suggests that 
velocity space structure may be generated primarily by nonlinear perpendicular phase 
mixing (instead of linear, parallel phase mixing).

Having chosen appropriate
indicators of velocity space resolution, we must devise a method for estimating the
error in these quantities.  This error depends on the particular numerical 
integration scheme used.  For
the energy and untrapped $\lambda$ integrals, which use Gaussian
quadrature, the error, $\epsilon_{G}$, is given by
\begin{equation}
\epsilon_{G} = \gamma_{m}f^{(2m)}(\zeta),
\label{eqn:gauss}
\end{equation}
where $f$ is the integrand, $m$ is the number of grid points, and $\zeta$ is some 
unknown point in the interval of integration.  The quantity $\gamma_{m}$ is
\begin{equation}
\gamma_{m} = \frac{2^{2m+1}\left(m!\right)^{4}}{\left(2m+1\right)\left[\left(2m\right)!\right]^{3}}
\label{eqn:gaussleg}
\end{equation}
for the untrapped $\lambda$ and finite domain energy integrals that use Gauss-Legendre
quadrature and
\begin{equation}
\gamma_{m} = \frac{\left(m!\right)^{2}}{\left(2m\right)!}
\label{eqn:gausslag}
\end{equation}
for the semi-infinite domain energy integral that uses Gauss-Laguerre quadrature.
The error, $\epsilon_{L}$, for the trapped $\lambda$ integrals, 
which use a newly upgraded integration scheme based on Lagrange interpolating 
polynomials (see e.g. Ref.~\onlinecite{hilde}), is given by
\begin{equation}
\epsilon_{L} = \frac{1}{m!}\int f^{(m)}(\zeta)  \pi(z)dz,
\label{eqn:lagrange}
\end{equation}
where
\begin{equation}
\pi(z)=\prod_{i=1}^{m}\left(z-z_{i}\right),
\end{equation}
with $z_{i}$ the $i^{th}$ grid point.  It should be noted that  $\zeta$ in Eqn 
(\ref{eqn:lagrange}) is an unknown function of $z$ whose domain is some subset
of the interval of integration.

From Eqns (\ref{eqn:gauss}) and (\ref{eqn:lagrange}), we see that Gaussian
quadrature gives exact results for polynomials of degree less than $2m$, while the 
Lagrangian method gives exact results only for polynomials of degree less than $m$. 
We say that the two schemes have degrees of precision $2m-1$ and $m-1$, respectively.
This difference arises because the
grid points in the Lagrangian method are fixed by boundary conditions, whereas the 
grid points in Gaussian quadrature are free parameters optimally chosen to 
improve the scheme's degree of precision.

Unfortunately, the formal error expressions ($\ref{eqn:gauss}$) and (\ref{eqn:lagrange})
are not very useful in practice:  they require information about high-order derivatives of
the distribution function, which is unavailable.  
As an alternative estimate for the error, we choose to compare multiple integral 
approximations computed with different degrees of precision, a common technique
in numerical analysis~\cite{zwillinger}.

\subsubsection{General description of the scheme}

Given the value of a function $f(z)$ at $N$ fixed points on the interval $[a,b]$, 
we would like to find two different approximations to the integral $\int_{a}^{b}f(z)dx$.
In our earlier discussion, we stated that an approximation with degree of precision $N-1$
can be found using a technique based on Lagrange interpolation; we call this approximation
$A_{h}$.  If we instead choose
to use only $M$ of the given functional values ($M<N$), we can use the same technique to 
find another integral approximation, $A_{l}$, with degree of precision $M-1$.  An 
estimate for the absolute error $\epsilon_{a}$ in the less accurate of these two 
approximations is obtained by taking the difference between the two:
\begin{equation}
\epsilon_{a} = \left|A_{h} - A_{l}\right|.
\end{equation}
Making the reasonable assumption that the approximation with higher degree of precision
is more accurate, $\epsilon_{a}$ represents the error in $A_{l}$.  However, it can
also be used as a more conservative error estimate for $A_{h}$.

If the $N$ points are chosen according to Gaussian quadrature rules, then one can find
an integral approximation with degree of precision $2N-1$.  As before, a second 
approximation can be obtained by using only $M$ of the $N$ grid points.  However,
due to the uniqueness of the grid points used for Gaussian quadrature, the $M$-point
grid no longer satisfies Gaussian quadrature rules.  As a result, this second 
approximation once again has degree of precision $M-1$.  Since the degrees of
precision of the two approximations differ by greater than a factor of two, the resulting
error estimate is likely to be very conservative when applied to $A_{h}$.  The factor
of approximately two difference in degree of precision makes this error estimate similar
to that obtained by comparing results from runs with $N$ and $N/2$ grid points, respectively
(for which the degrees of precision would be $2N-1$ and $N-1$).


The conservative nature of the error estimate for $A_{h}$ depends upon our assumption 
that a higher degree of precision results in a more accurate integral approximation.  
For Gaussian quadrature, it can be shown that the error in the integral approximation can be
made arbitrarily small by choosing the degree of precision large enough~\cite{hilde}.  
The same result
does not necessarily hold for the Lagrangian method with arbitrary grid spacing because
the weights in this case are not all guaranteed to be positive.  However, the 
error $\epsilon_{M}$ in an $M$-point integral approximation satisfies
\begin{eqnarray}
\epsilon_{M} &\leq 2& \epsilon \sum_{i=1}^{M}\left|w_{i}^{(M)}\right| \\
&\leq& 2 \epsilon M \max_{i=1,M} \left|w_{i}\right| \\
&=& 2 \epsilon M \kappa(M),
\end{eqnarray}
where $\epsilon$ can be chosen arbitrarily small for large enough $M$, and $w_{i}^{(M)}$
is the weight corresponding to the $i^{th}$ grid point out of $M$.  From this result,
we see that as long as $\kappa$ is bounded when $M\rightarrow\infty$, then 
$\epsilon_{M}\rightarrow0$ as $M\rightarrow \infty$.  This cannot be verified in
advance, but one can gain confidence by checking a posteriori.  In practice, we calculate
$\kappa$ for the chosen $M$ and subdivide the integration domain into subintervals
with fewer points if $\kappa$ is larger than some reasonable value.
  
 \subsubsection{Implementation in \texttt{GS2}} 
  
  In \texttt{GS2}, we must compute two-dimensional integrals over energy and $\lambda$.  As
  stated in Sec. III, each of these integrals is effectively separated into two by splitting
 the $\lambda$ integration into trapped and untrapped regions.  Since the number of grid
 points in energy and both $\lambda$ regions can be varied independently of each other,
 we wish to monitor resolution in each of these three variables individually.  This entails
 computing three separate integral
error estimates: one for energy integrals, one for
 untrapped $\lambda$ integrals, and one for trapped $\lambda$ integrals.
 
These integral error estimates are calculated using the technique described in the previous
subsection.  For energy and untrapped $\lambda$ integrals, Gaussian quadrature is used
to obtain the two-dimensional integral approximation $A_{h}$.  This approximation
has degree of precision $2N_{E}-1$ for the energy integration and $2N_{u}-1$
for the untrapped $\lambda$ integration, where $N_{E}$ and $N_{u}$ are the number 
of energy and untrapped $\lambda$ grid points, respectively.  To obtain the second
approximation, $A_{l}$, we fix the grid and weights for one variable and drop one grid point for the other variable,
recomputing the weights.  As an example, we choose to drop an untrapped $\lambda$ grid 
point.  The degree of precision for $A_{l}$ is then $2N_{E}-1$ for the energy integration
and $N_{u}-2$ for the untrapped $\lambda$ integration.
Since there is nothing special about the particular grid point we drop, we repeat the process a
total of $N_{u}$ times, each time dropping a different point and computing a different set of 
weights.  The final error estimate is an average of these error estimates.

For the trapped $\lambda$ integrals, Lagrangian quadrature is used to obtain $A_{h}$,
which has degree of precision $N_{t}-1$.  We obtain the approximation $A_{l}$
by dropping two points symmetrically about $v_{\parallel}=0$, as shown in 
Fig.~\ref{fig:GS2v}.  We drop an additional point here because it provides a
slightly more conservative error estimate and because maintaining the symmetry of the 
grid points provides better stability for the weights associated with the Lagrange 
interpolation scheme.  As before, we repeat this process for each possible grid point 
pair and take the average of the individual error estimates to get the final error estimate.
  
All modified grids and weights necessary for the integral error estimates are computed once
at initialization and need not be computed again.  The additional integrations necessary to 
obtain our error estimates are computationally
cheap when compared to the expense of solving for the distribution function and fields
at each time step.  Furthermore, we do not need an error estimate at each time step, 
so the diagnostic can be used sparingly.  Consequently, our error estimate comes
at essentially no extra cost.

\subsection{Spectral method}

An alternative method for testing v-space resolution is to expand the velocity space 
distribution function in an appropriate basis set and monitor the amplitude of the basis
function coefficients.  Whenever the highest mode number coefficients that can be
accurately calculated in the simulation acquire appreciable
amplitudes, we can no longer feel confident that the simulation is resolved.  
Since we choose our grid points according to Gauss-Legendre
quadrature, it is convenient (and most accurate) to choose the Legendre polynomials as 
our basis functions.  The coefficient of the $m^{th}$ Legendre polynomial in the
expansion of $h$ is given by
\begin{eqnarray}
c_{m} &=& \frac{2m+1}{2}\int_{-1}^{1} h(s) P_{m}(s) ds \\
&\approx& \frac{2m+1}{2} \sum_{i=1}^{n_{\epsilon}-1} w_{i} h(s_{i}) P_{m}(s_{i}),
\label{eqn:cm}
\end{eqnarray}
where $P_{m}$ is the $m^{th}$ Legendre polynomial, and $\{w_{i}\}$ are the weights
associated with Gauss-Legendre quadrature.
The integral approximation in Eqn.~(\ref{eqn:cm}) has degree of precision $2N-1$.  
Assuming $h$ has a degree of at least $m$ (otherwise $c_{m}=0$), our approximation
for $c_{m}$ is only exact for $m<N$.

There are various ways in which one could use these $\{c_{m}\}$ to estimate the error
in velocity space resolution.  We assume locality of interaction between the various 
modes so that we only have to monitor the amplitudes of the few highest modes.  At each 
($\theta$, $k_{x}$, $k_{y}$)-point, we find the maximum amplitude of the three highest
mode number spectral coefficients, $c_{h,max}$,
and the maximum amplitude of all the spectral coefficients, $c_{max}$.  We then use
the following normalized sum as a relative estimate for the error:
\begin{equation}
\epsilon_{c} = \sum_{\theta,k_{x},k_{y}} c_{h,max}(\theta,k_{x},k_{y})/
\sum_{\theta,k_{x},k_{y}}c_{max}(\theta,k_{x},k_{y}).
\label{eqn:specErr}
\end{equation}
When the normalized amplitude $\epsilon_{c}$ grows too large, we can no
longer be confident that the simulation is resolved.  Of course, how large $\epsilon_{c}$
can get before resolution suffers varies from problem to problem.  As before with
the integral method, we determine a scaled estimate of the error based on empirical
evidence from a wide range of simulation data.

\subsubsection{Application of error diagnostics}

We have applied both the integral and spectral error diagnostics to a diverse set of 
simulations, including:
linearly growing modes such as the electron drift wave and the ITG mode; 
linearly damped modes such as the ion acoustic wave and kinetic Alfven wave;
neoclassical transport;
and nonlinear dynamics of slab ETG and toroidal ITG modes.  From these simulations, 
we have determined empirical scaling factors for our conservative error estimates.  Here,
we present typical results from a cross-section of the above simulations.

\begin{figure}
\centering
\includegraphics[height=3.0in]{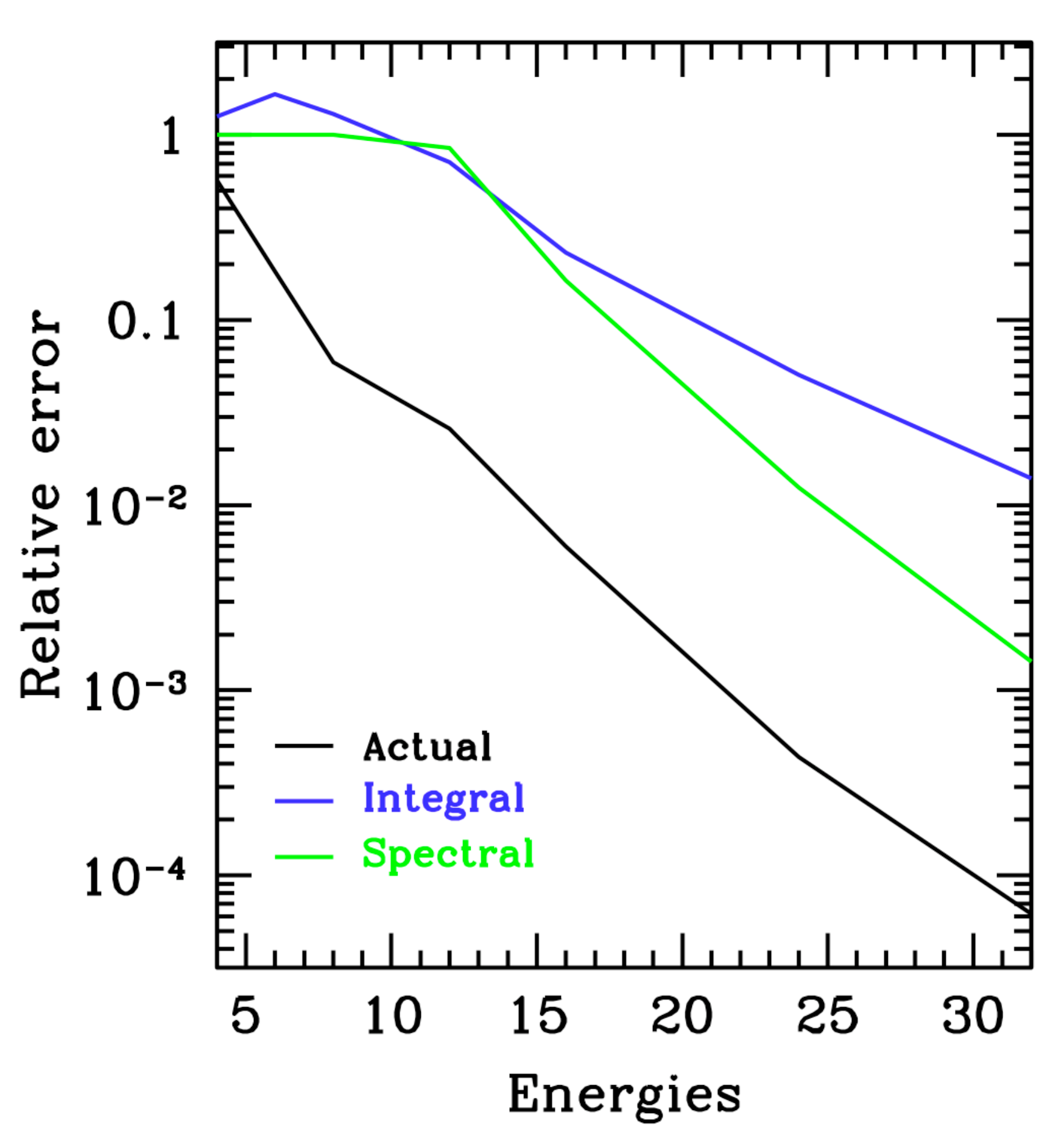}
\includegraphics[height=3.0in]{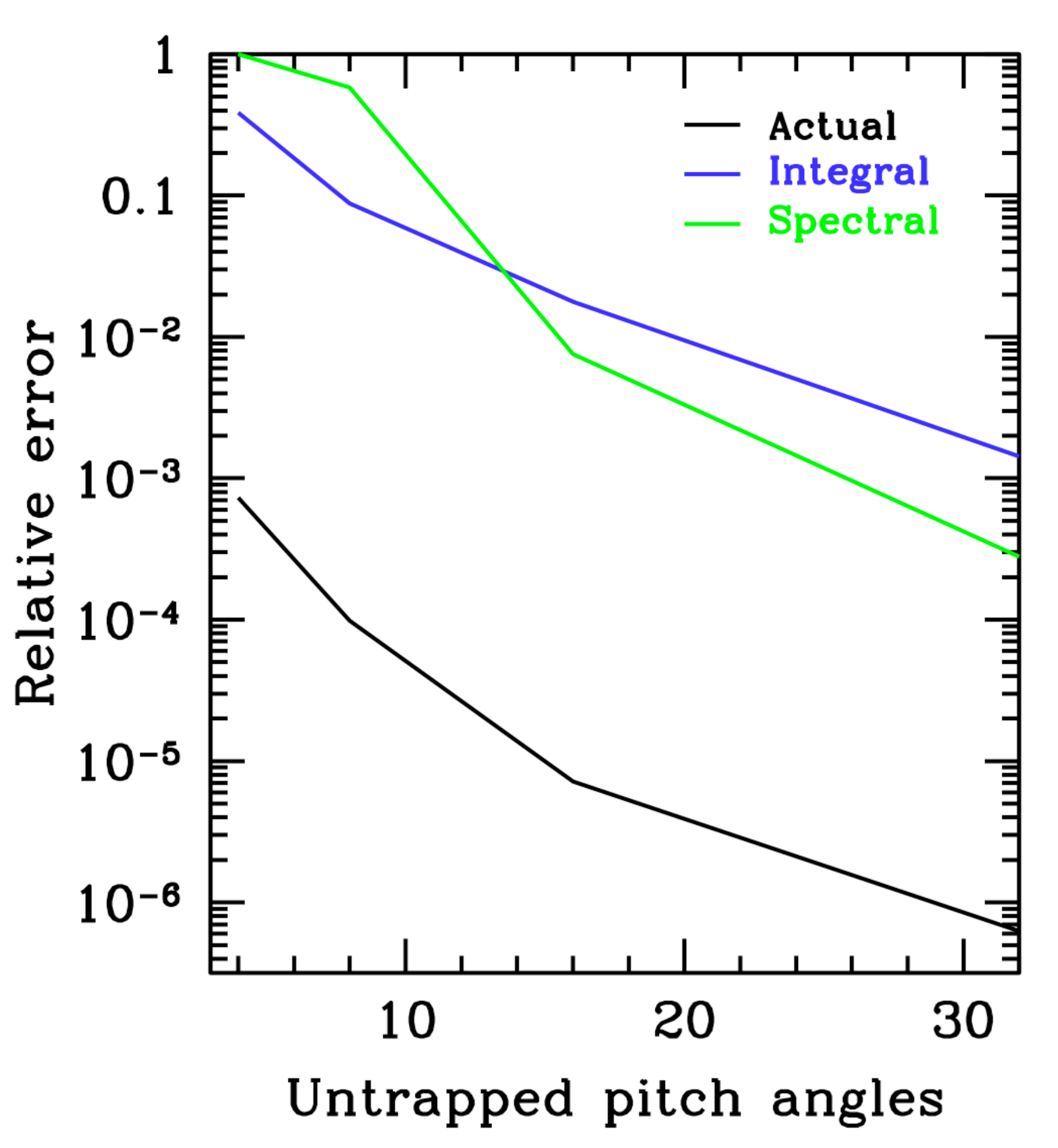}
\caption{Comparison of actual and (unscaled) estimated error in wave frequency due to insufficient
resolution in energy (left) and untrapped $\lambda$ (right).
The actual wave frequency, $\omega$, is determined
from a higher resolution run with 64 grid points in energy and both trapped and untrapped
$\lambda$.  The actual relative error, $\epsilon$, is then defined to be $\epsilon=
\sqrt{\frac{\left|\omega-\omega_{n}\right|^{2}}{\left|\omega\right|^{2}}}$,
where $\omega_{n}$ is the approximation to $\omega$ obtained from a run with $n$
grid points.}
\label{fig:lincycerr}
\end{figure}

\begin{figure}
\centering
\includegraphics[height=5.0in]{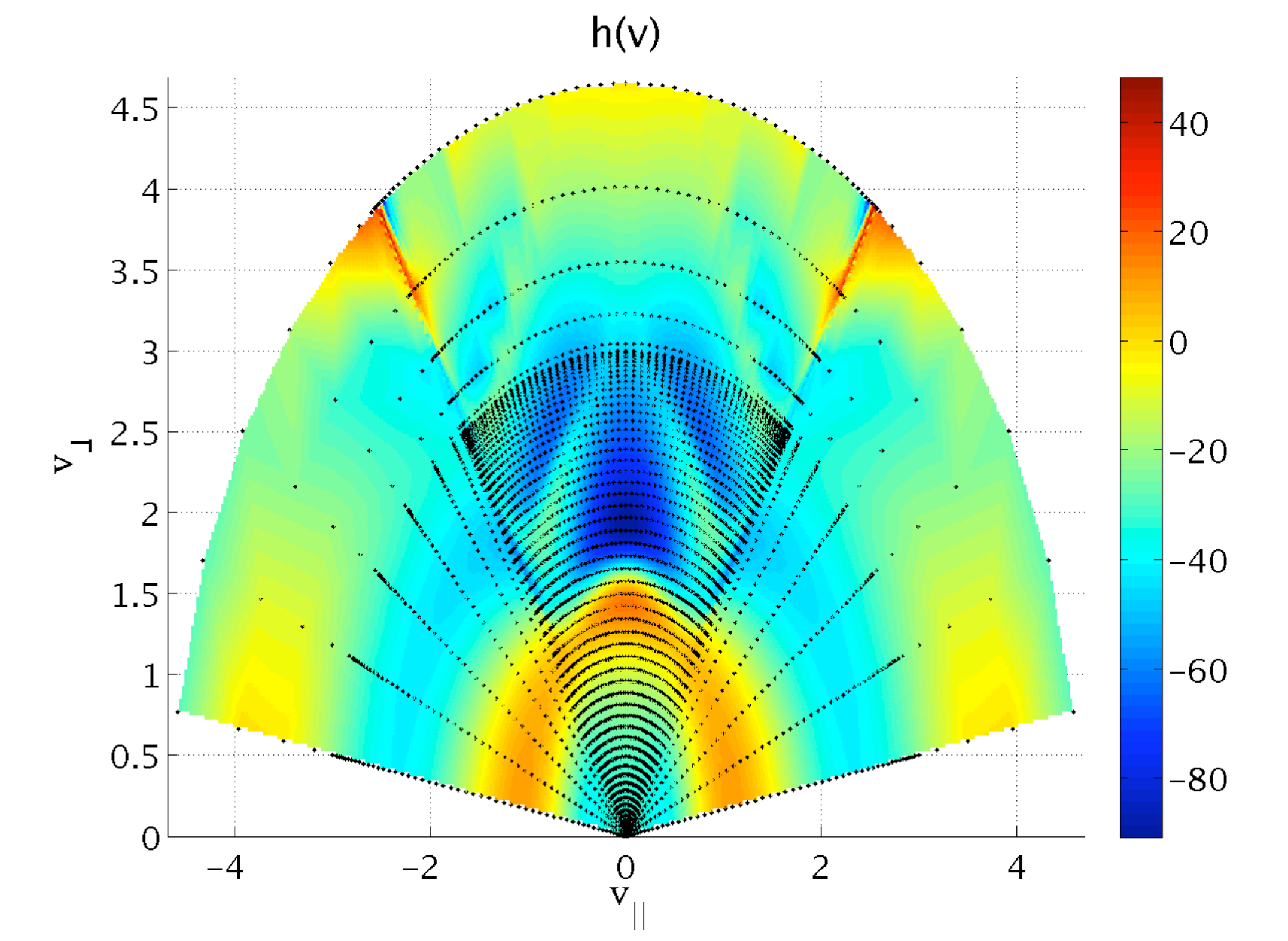}
\caption{Non-Boltzmann part of the perturbed distribution function, normalized by
$F_{0}$, for the linear, toroidal ITG mode with Cyclone base-case parameters.  The use of a polar grid in velocity space, as well as the fine mesh near the trapped-passing boundary, minimizes the number of
grid points necessary for resolution.}
\label{fig:lincycdist}
\end{figure}

Fig.~\ref{fig:lincycerr} compares the unscaled error estimates 
in energy and $\lambda$ with the actual errors in
growth rate as we vary the number of grid points in a linear simulation of the collisionless
toroidal ITG mode (using Cyclone base case parameters~\cite{dimitsPoP00}).  
The simulation remains well-resolved down to very few 
grid points, and the error estimates agree well with the 
actual error.  The error due to resolution in untrapped $\lambda$ is still small for as little as
four grid points due to our choice of velocity variables, as illustrated by the snapshot
of the distribution
function shown in Fig.~\ref{fig:lincycdist}.

\begin{figure}
\centering
\includegraphics[height=3.0in]{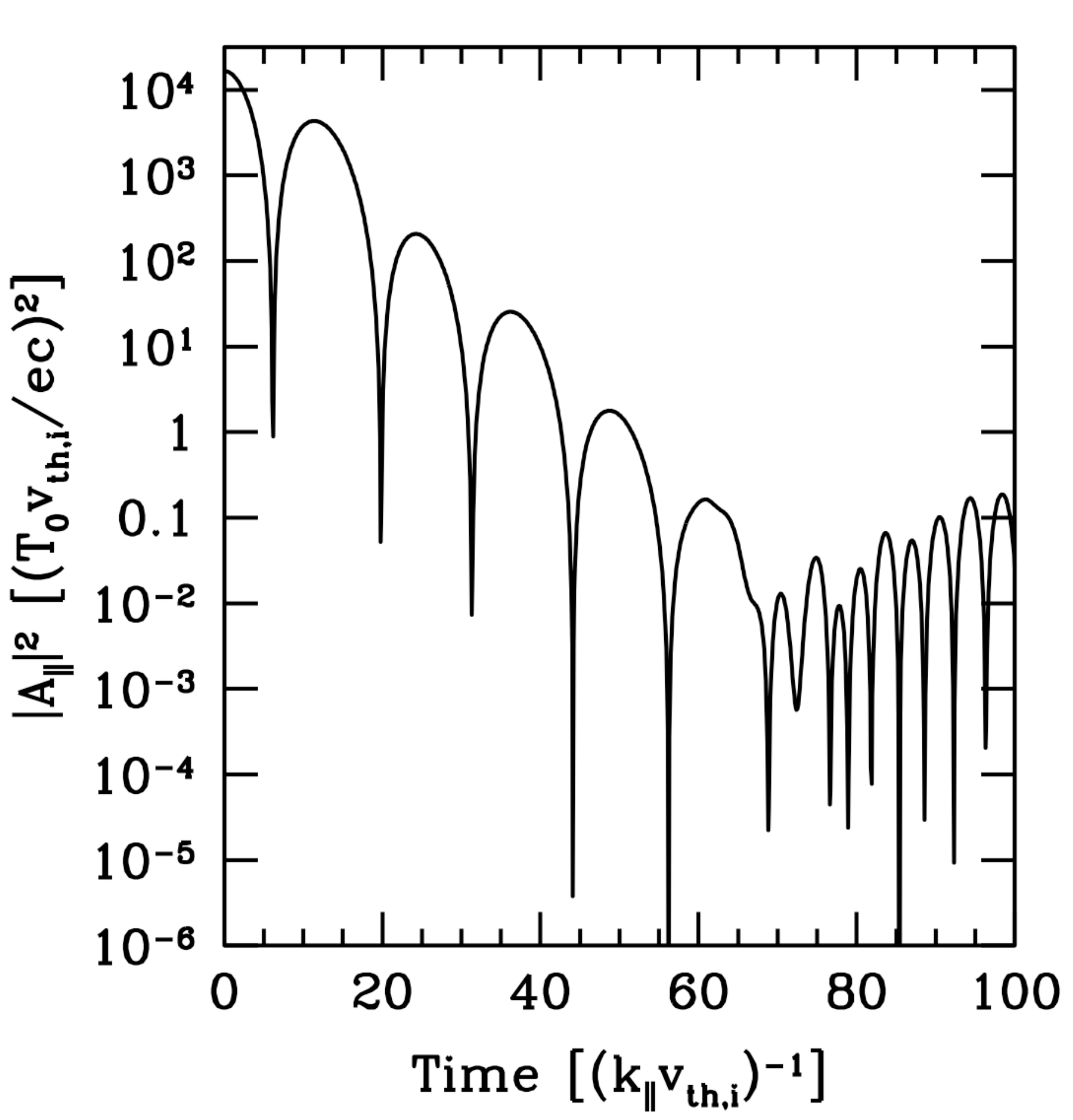}
\includegraphics[height=3.0in]{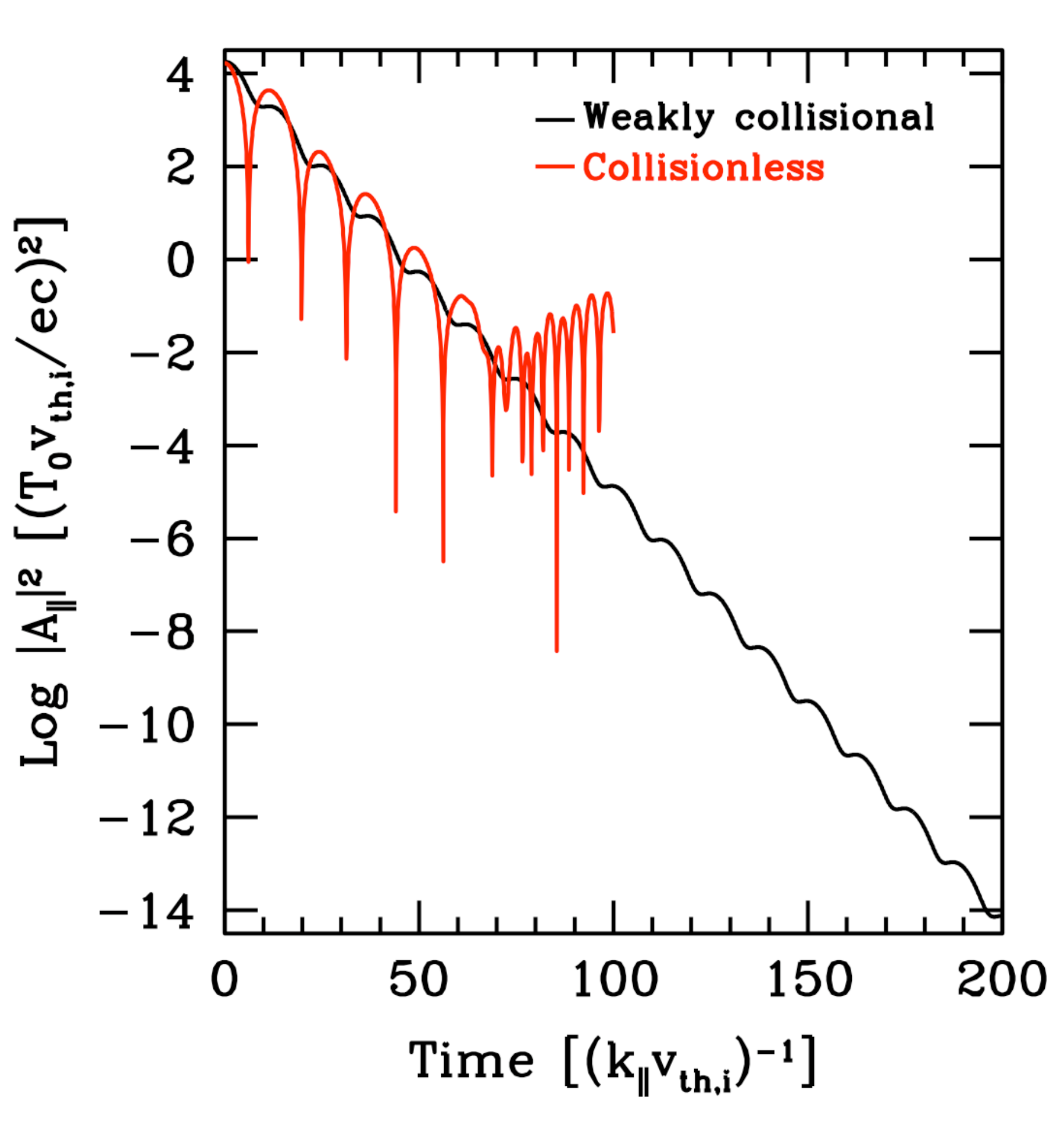}
\caption{Barnes damping of the kinetic Alfven wave for simulations with 16 energy grid points and 32 pitch angles for each sign of $v_{\parl}$.  In the absence of collisions (left),
sub-grid scale structures develop in velocity space, and the damping is artificially terminated.
A small collisionality ($\nu \ll \gamma$) prevents the development of
sub-grid scale structures in velocity space, and the damping rate remains correct
indefinitely (right).}
\label{fig:kawdamp}
\end{figure}

\begin{figure}
\centering
\includegraphics[height=3.2in]{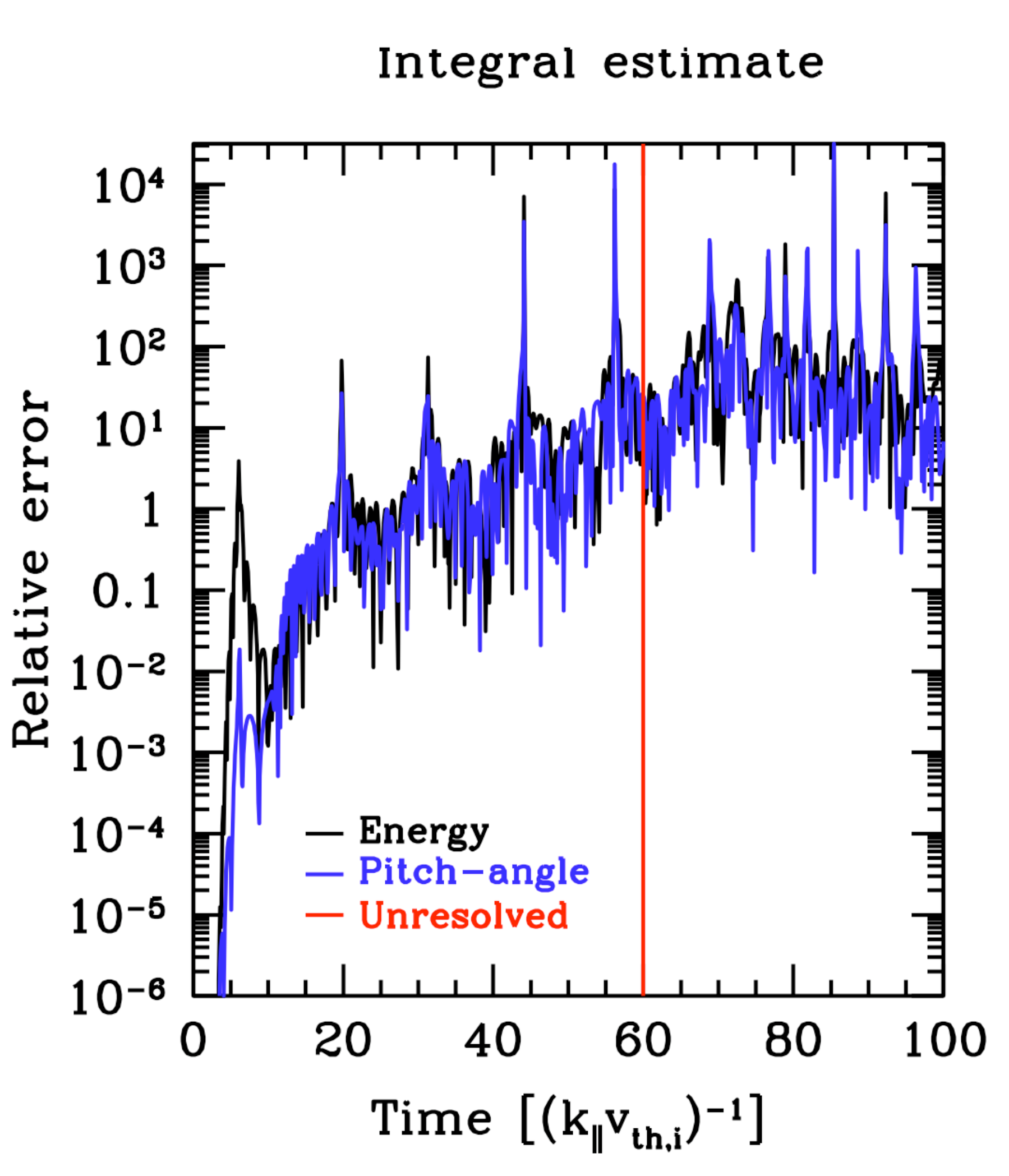}
\includegraphics[height=3.2in]{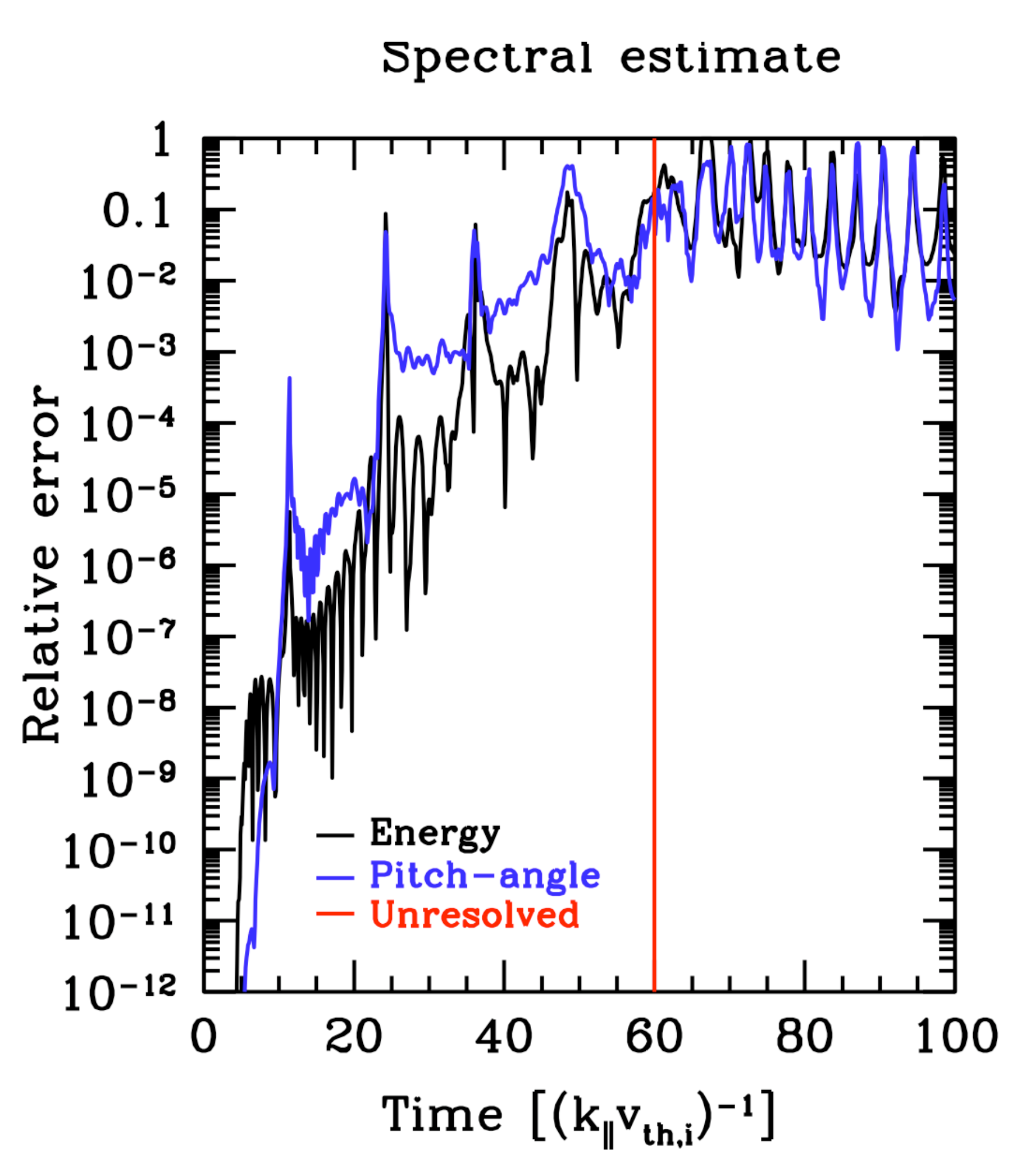}
\caption{Time evolution of the integral (left) and spectral (right) error estimates for the collisionless kinetic Alfven wave.  Vertical line represents time at which damping rate artificially terminates due to poor resolution.}
\label{fig:kawerr}
\end{figure}

\begin{figure}
\centering
\includegraphics[height=3.2in]{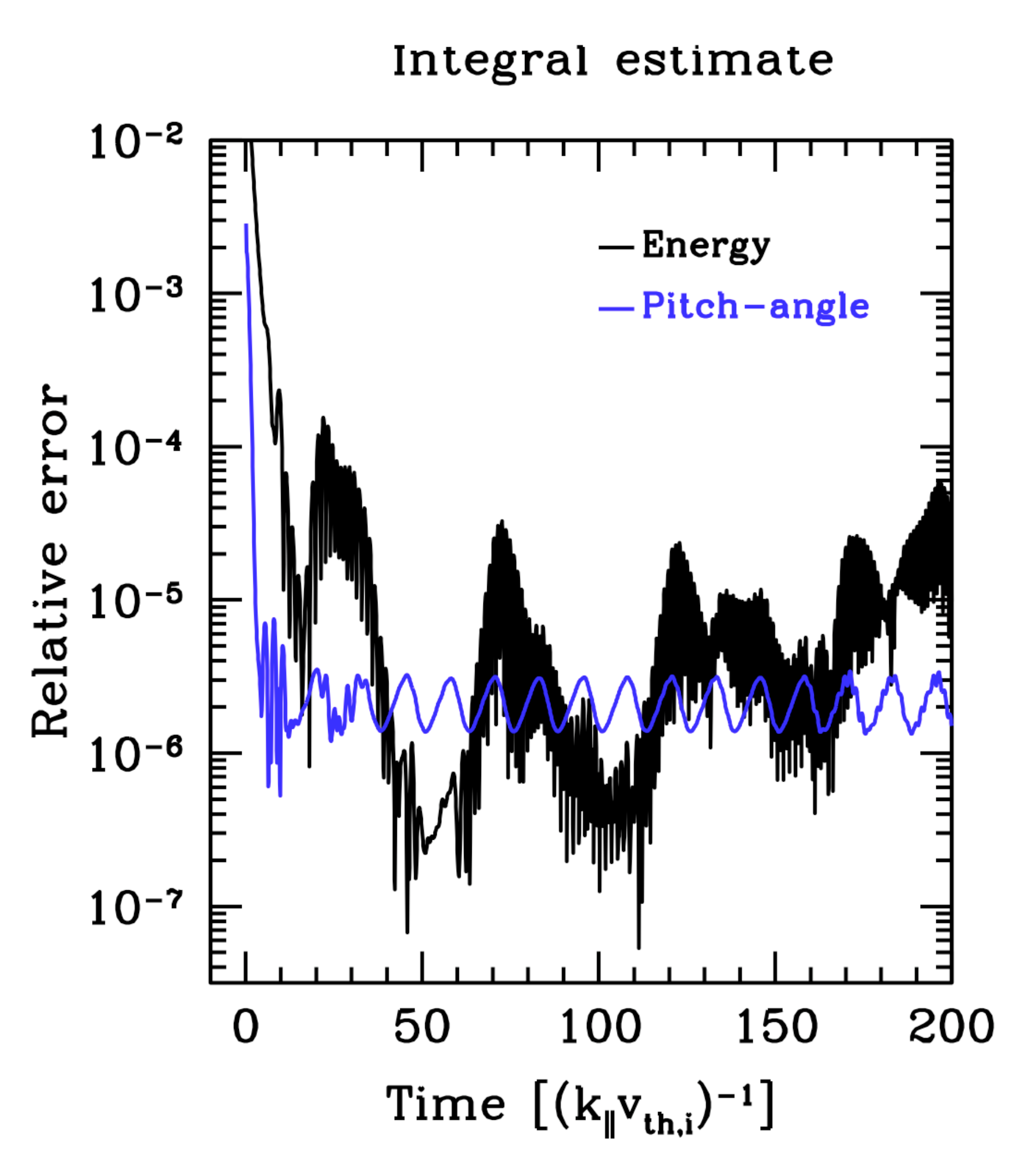}
\includegraphics[height=3.2in]{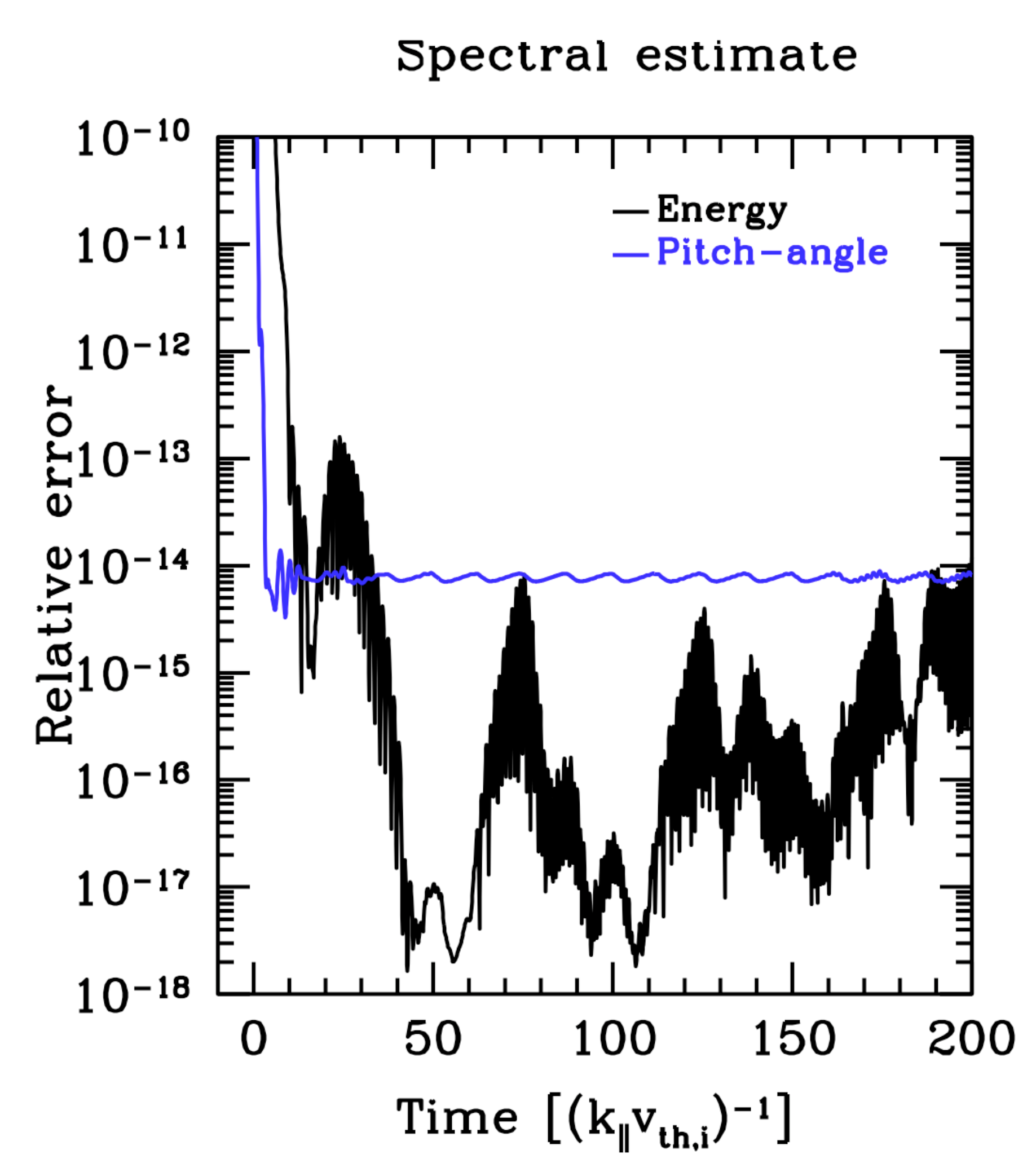}
\caption{Time evolution of integral and spectral error estimates for the weakly collisional kinetic Alfven wave damping.  The estimates correctly indicate that the simulation remains well resolved indefinitely.}
\label{fig:kawerrcoll}
\end{figure}

Figs.~\ref{fig:kawdamp} and~\ref{fig:kawerr} show the damping of $A_{\parl}$ and
the corresponding scaled error estimates for the simulation of a collisionless 
kinetic Alfven wave with 16 energy grid points and 32 pitch angles for each sign of the parallel velocity. 
The collisionless damping rate in Fig.~\ref{fig:kawdamp} agrees with theory until 
sub-gridscale structure develops in velocity space, at which point damping ceases.  The
onset of sub-gridscale structure corresponds to the peak in scaled error in 
Fig.~\ref{fig:kawerr}.  The addition of a small 
collisionality prevents sub-gridscale structure, as shown in Fig.~\ref{fig:kawdamp}, 
where the damping rate of $A_{\parl}$
agrees well with theory indefinitely.  This is accurately predicted by the error estimates of 
Fig.~\ref{fig:kawerrcoll}, which never reach appreciable magnitude.

\section{Adaptive collision frequency}
\label{sec:adaptC}

As stated earlier, we would like to know what combination of dissipation and grid spacing
is necessary for a resolved simulation.  One way to approach this problem is
to fix the dissipation and vary the number of grid points to find how many are required
to get an accurate result.  This is the general idea behind the error estimation diagnostics
described in the previous section.  However, if we wanted to use this approach to ensure
that the simulation remained resolved, we would have to implement an adaptive grid,
which is difficult to do on massive, multi-processor machines.  

Instead, we choose an alternative approach: we fix the number of grid points and vary
the dissipation until we have a well-resolved result.  In particular, we have implemented
an adaptive collision frequency in \texttt{GS2} that allows for the independent variation of 
the collisionality associated with pitch-angle scattering and energy diffusion.  Given
an acceptable error tolerance for velocity space calculations, a scaled version of the 
integral error estimate described in the previous section is used to determine whether
or not the simulation is well-resolved.  The collision frequency is then adjusted using
a feedback process until the scaled estimate of the error converges to within some
pre-specified window of the desired error tolerance.  In this way, the approximate 
minimum possible
dissipation is used to achieve an acceptable degree of resolution in velocity space.

Of course, the amount of dissipation necessary to resolve a simulation at a fixed number
of grid points may be quite large if a coarse grid is used.  Consequently, the collisionless
dynamics may be modified.  As a result, it is necessary to compare the converged 
collision frequency with dynamic frequencies of interest in the problem.

\begin{figure}
\centering
\includegraphics[height=3.0in]{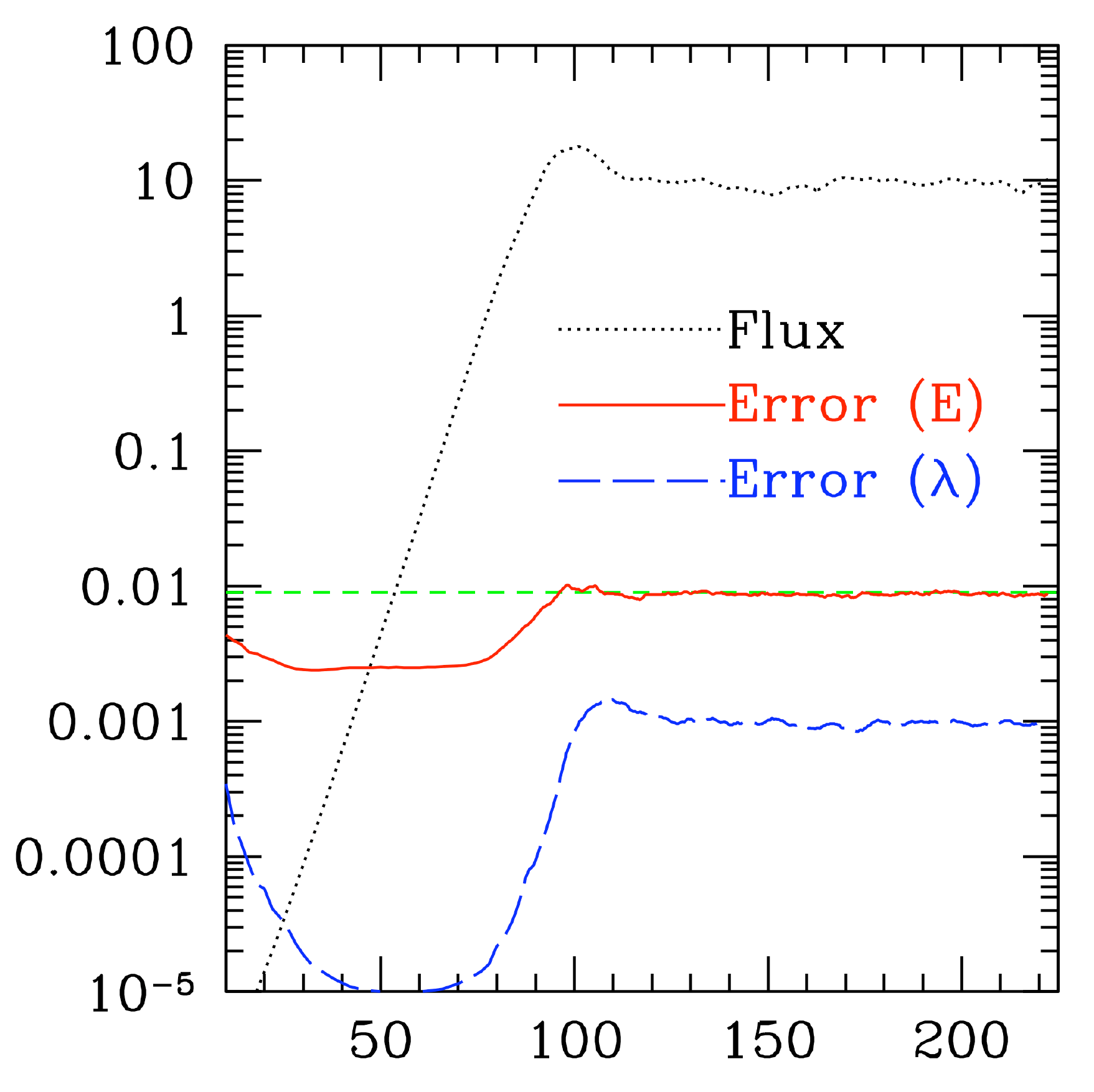}
\includegraphics[height=3.0in]{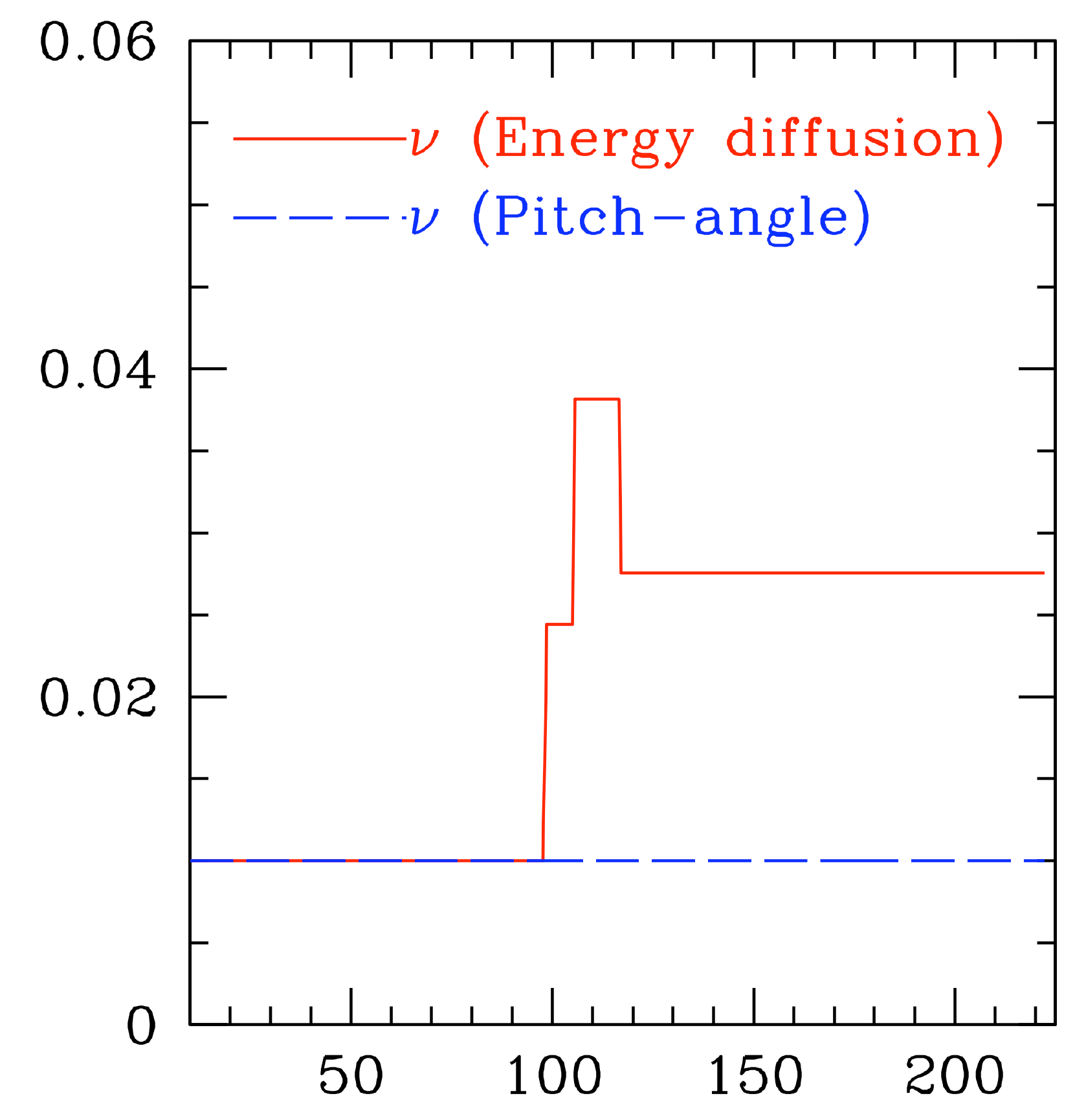}
\caption{(Left): Normalized electron heat flux vs. time for a nonlinear simulation
of ETG turbulence.  Scaled estimates of the error in energy and $\lambda$ resolution
increase during nonlinear saturation, but are kept within the specified relative error tolerance of $0.01$
with the use of an adaptive collision frequency.  (Right): Collision frequency (normalized 
by $k_{\parl}v_{th,e}$) vs. time.}
\label{fig:adaptetg}
\end{figure}

As an example we consider a nonlinear simulation of electron temperature gradient (ETG)
turbulence in slab geometry (i.e. straight background magnetic field).  In the nonlinear
phase, small scales are expected to develop in velocity space, potentially challenging numerical 
resolution.  In Fig.~\ref{fig:adaptetg}, we see that this is indeed the case.  Our velocity
space resolution diagnostics indicate that the errors in velocity space begin to increase
sharply during the transition from linear instability to turbulence.  However, our use of
an adaptive collision frequency prevents the estimated error from exceeding the user-defined
relative error tolerance (in this case, $0.01$).  We see that the error remains on the threshold
of the error tolerance, while the collision frequency for energy diffusion increases to a 
steady-state value of $\nu\approx0.027 \ k_{\parl}v_{th,e}$, which is well below the
dynamic frequency in the system.  Consequently, the collisionless dynamics are unaltered.

\section{Summary}
\label{sec:vspacesum}

In this paper, we discussed the development of small-scale structure in velocity space,
presented a set of velocity space resolution diagnostics for use in gyrokinetic simulations,
and introduced an adaptive collisionality that allows us to resolve simulations with
an approximate minimal necessary dissipation for a fixed number of grid points in velocity space.  
In \secref{sec:gkv}
we demonstrated the tendency of collisionless plasmas to develop increasingly fine
scales in the distribution of particle velocities and discussed the phase mixing processes
that lead to such behavior.

In \secref{sec:gs2v} we described the treatment of velocity space in the gyrokinetic
code \verb#GS2#.  We gave details on the choice of velocity space variables
(energy and pitch-angle) and discretization scheme, which is chosen to minimize
the error of the numerical integrals necessary to obtain the electromagnetic fields.  
This included presentation of a newly implemented energy grid, which provides spectrally 
accurate integrals over particle energies.  Additionally, we gave a brief discussion of
both the physical and numerical dissipation mechanisms available for use in \verb#GS2#.

We discussed common approaches to monitoring velocity space resolution in \secref{sec:vresdiag}
and the difficulties associated with each.  We then proposed two new measures of velocity
space resolution and detailed implementation in \verb#GS2#.  One of the proposed
resolution diagnostics involves obtaining estimates for the error in field integrals by
comparing numerical integrals obtained using integration schemes with differing degrees
of precision.  The other resolution diagnostic involves decomposing the perturbed distribution
function into spectral components in velocity space and monitoring the amplitude of the
spectral coefficients.  Both diagnostics should be quite conservative.

We then applied our resolution diagnostics to a number of example problems, including
Landau damping of the ion acoustic wave, Barnes damping of the kinetic Alfven wave,
and linear instability of the toroidal ITG mode.  We found that both diagnostics do well
in qualitatively estimating errors due to limited velocity space resolution.  Due
to their conservative nature, an empirical scaling factor was necessary to obtain correct
quantitative predictions.  

In \secref{sec:adaptC} we coupled the error estimates from our resolution diagnostics
with a model physical collision operator to develop an adaptive collision frequency.  This
adaptive collision frequency allowed us to resolve velocity space while using an approximate
minimal necessary amount of dissipation.  When using the adaptive collision frequency, one
must monitor the ratio of the collision frequency to the dynamic frequency to ensure
that one is still within the weakly collisional regime.

In conclusion, we found that dissipation was not necessary to resolve linear 
instabilities, but it was necessary to resolve nonlinear dynamics and linearly damped waves.  
For the nonlinear cases considered here (slab ETG and toroidal ITG), the required 
collisionality for resolution obtained with the adaptive collision frequency was
found to be no larger than the physical collisionality used in modern fusion experiments.

\appendix
\section{Landau-damped ion acoustic wave}

We consider the collisionless ion acoustic wave in slab geometry with adiabatic electrons.
The gyrokinetic equation for this system has the particularly simple form
\begin{equation}
\pd{h}{t} + v_{z}\pd{h}{z} = \frac{qF_{0}}{T}\pd{\lb \Phi \rb}{t}.
\label{eqn:iaw}
\end{equation}
Changing variables from $h$ to $g\equiv\lb f_{1} \rb$ and assuming solutions of the form
\begin{equation}
g = \tilde{g}(\mbf{v})e^{i\left(k_{\parl}z - \omega t\right)},
\end{equation}
we obtain
\begin{equation}
\left( \omega - kv\right) g = kv \frac{e \lb \Phi \rb_{\mbf{R}}}{T_{i}} F_{M},
\end{equation}
where we are using $v=v_{\parl}$ and $k=k_{\parl}$ for convenience. 
Neglecting FLR effects and assuming quasineutrality gives
\begin{equation}
\left(\omega - kv\right)g = kv \tau \frac{F_{M}}{n_{0}}\int d^{3}v' g(v').
\label{depart}
\end{equation}
Defining
\begin{equation}
\overline{g}(v) = 2\pi \int_{0}^{\infty}v_{\perp} dv_{\perp} g(\mbf{v})
\end{equation}
and integrating over the perpendicular velocities in the gyrokinetic equation yields
\begin{equation}
\left(\omega-kv\right) \overline{g}(v) = k F(v),
\label{start}
\end{equation}
where
\begin{eqnarray}
F(v) &=& v \tau \frac{n_{1}}{\sqrt{2\pi}v_{th}}e^{-\frac{v^{2}}{2v_{th}^{2}}},\\
n_{1} &=& \int dv' \overline{g}(v').
\end{eqnarray}
Following the analysis of Refs.~\onlinecite{vankamp55} and~\onlinecite{case59}, we see that this equation has 
solutions of the form
\begin{eqnarray}
\overline{g}(v) &=& F(v)\left[\mathcal{P}\frac{1}{u-v}+\Lambda(k,u) \delta(u-v)\right], \label{g}
\end{eqnarray}
with $u=\frac{\omega}{k}$, provided that $\Lambda$ is chosen to satisfy the condition
\begin{equation}
n_{1} = \int dv' \overline{g}(v') = \mathcal{P}\int dv' \frac{F(v')}{u-v'} + \Lambda(k,u) F(u).
\label{lamcon}
\end{equation}
A general solution is given in the form
\begin{equation}
\overline{f}(z,v,t) = \int_{-\infty}^{\infty} \int_{-\infty}^{\infty} \mathcal{C}(k,u) \overline{g}_{k,u}(v) e^{ik\left(z-ut\right)}dk \ du,
\label{gensol}
\end{equation}
where $\mathcal{C}(k,u)$ is determined by the initial condtion
\begin{equation}
\overline{f}(z,v,0) = \int \int \mathcal{C}(k,u) \overline{g}_{k,u}(v)e^{ikz}dk \ du.
\end{equation}

Taking the inverse Fourier transform of the above expression gives
\begin{equation}
\mathcal{F}(k,v) = \int \mathcal{C}(k,u) \overline{g}_{k,u}(v) du,
\label{ic}
\end{equation}
where
\begin{equation}
\mathcal{F}(k,v) = \frac{1}{2\pi}\int \overline{f}(z,v,0) e^{-ikz} dz.
\end{equation}
Plugging the expression (\ref{g}) for $\overline{g}$ into the initial condition (\ref{ic}) yields
\begin{equation}
\frac{\mathcal{F}(k,v)}{F(v)} = \mathcal{P}\int\frac{\mathcal{C}(k,u)}{u-v}du + \Lambda(k,v) \mathcal{C}(v).
\label{ccon}
\end{equation}

We now have two equations, (\ref{lamcon}) and (\ref{ccon}), for two unknowns ($\Lambda$ and $\mathcal{C}$).  In order to solve
this linear system, it is convenient to define some new notation.  Any square integrable function $H$ can be written
\begin{equation}
H(q) = \int_{-\infty}^{\infty} K(p) e^{ipq} dp. 
\end{equation} 
We define the positive and negative frequency parts of $H$ as
\begin{eqnarray}
H_{\pm}(q) &=& \pm \int_{0}^{\pm\infty} K(p) e^{ipq} dp,
\end{eqnarray}
so that $H = H_{+} + H_{-}$. Further we define the function $H_{*}=H_{+}-H_{-}$.  It 
can be shown that $H_{*}$ has the alternate form
\begin{equation}
H_{*}(v) =  \mathcal{P} \frac{1}{\pi i}\int_{-\infty}^{\infty} \frac{H(v')}{v'-v}dv'.
\end{equation}
With these definitions in hand, we rewrite eqns (\ref{lamcon}) and (\ref{ccon}) as
\begin{eqnarray}
n_{1} &=& -\pi i F_{*}(u) + \Lambda F(u),\\
\frac{\mathcal{F}(k,v)}{F(v)} &=& \left(\Lambda + \pi i\right)\mathcal{C}_{+}(v) + \left(\Lambda - \pi i\right)\mathcal{C}_{-}(v).
\end{eqnarray}
Eliminating $\Lambda$ gives an expression involving $\mathcal{C}_{+}$ and $\mathcal{C}_{-}$:
\begin{equation}
\mathcal{F}(k,u) = \left(n_{1} + 2\pi i F_{+}(u)\right) \mathcal{C}_{+}(k,u) + \left(n_{1} - 2\pi i F_{-}(u)\right)\mathcal{C}_{-}(k,u). 
\end{equation}
The transform $\mathcal{F}$ can also be broken down into negative and positive frequency parts to give two separate equations.
\begin{eqnarray}
\mathcal{F}_{\pm}(k,u) &=& \left(n_{1} \pm 2\pi i F_{\pm}(u)\right)\mathcal{C}_{\pm}(u)
\end{eqnarray}
These can then be used to construct $\mathcal{C}(k,u)$:
\begin{equation}
\mathcal{C}(k,u) = \frac{\mathcal{F}_{+}(k,u)}{n_{1}+2\pi i F_{+}(u)} + \frac{\mathcal{F}_{-}(k,u)}{n_{1}-2\pi i F_{-}(u)}.
\label{c}
\end{equation}

Substituting the expressions $(\ref{g})$ and $(\ref{c})$ for $\overline{g}$ and $\mathcal{C}$ into the equation (\ref{gensol}) for $\overline{f}$
gives
\begin{equation}
\begin{split}
\overline{f}(z,v,t)=\int \int &\left[\frac{\mathcal{F}_{+}(k,u)}{n_{1}+2\pi i F_{+}(u)}+\frac{\mathcal{F}_{-}(k,u)}{n_{1}-2\pi i F_{-}(u)}\right]F(v)\\
&\times \left[\mathcal{P}\frac{1}{u-v} + \Lambda(k,u)\delta(u-v)\right]e^{ik\left(z-ut\right)}dk \ du.
\label{f1}
\end{split}
\end{equation}
We can use the identity
\begin{equation}
\mathcal{F}_{\pm}(k,u) = \frac{1}{2\pi}\int_{-\infty}^{\infty} e^{-ikz'} dz' \int_{-\infty}^{\infty} \delta_{\pm}(u-v') \overline{f}(z',v',0) dv'
\end{equation}
to rewrite eqn (\ref{f1}) in the more convenient form
\begin{equation}
\begin{split}
\overline{f}(z,v,t) = \int &\left[\frac{\delta_{+}(u-v')}{n_{1}+2\pi i F_{+}(u)}+\frac{\delta_{-}(u-v')}{n_{1}-2\pi i F_{-}(u)}\right]\frac{\overline{f}(z',v',0)}{2\pi}F(v)\\
&\times\left[\mathcal{P}\frac{1}{u-v} + \Lambda(k,u)\delta(u-v)\right]e^{ik\left(z-z'-ut\right)}dz' dv' dk \ du.
\end{split}
\end{equation}
Now we pick an initial condition of the form
\begin{eqnarray}
\overline{f}(z,v,0) &=&\tilde{f}(v,0) e^{ik_{0}z},
\end{eqnarray}
which gives
\begin{eqnarray}
\overline{f}(z,v,t) &=& e^{ik_{0}\left(z-vt\right)}\left(n_{1}+\pi i F_{*}(v)\right)\left(\frac{\tilde{f}_{+}(v,0)}{n_{1}+2\pi i F_{+}(v)}+\frac{\tilde{f}_{-}(v,0)}{n_{1}-2\pi i F_{-}(v)}\right)\\
&+& \mathcal{P}\int \frac{F(v)}{u-v}\left(\frac{\tilde{f}_{+}(u,0)}{n_{1}+2\pi i F_{+}(u)}+\frac{\tilde{f}_{-}(u,0)}{n_{1}-2\pi i F_{-}(u)}\right)e^{ik_{0}\left(z-ut\right)} du.
\end{eqnarray}

\end{document}